\documentclass[fleqn,usenatbib]{mnras}

\usepackage{newtxtext,newtxmath}

\usepackage[T1]{fontenc}

\DeclareRobustCommand{\VAN}[3]{#2}
\let\VANthebibliography\thebibliography
\def\thebibliography{\DeclareRobustCommand{\VAN}[3]{##3}\VANthebibliography}

\usepackage{graphicx}	
\usepackage{amsmath}	
\usepackage{esvect}
\usepackage{fontawesome}

\title[Sky Motion of Interstellar Objects]{Sky-Plane Velocity Distributions of Interstellar Objects and Implications for Their Detection}

\author[Walker $\&$ Seligman]{
Cassidy E. Walker,$^{1}$\thanks{E-mail: cwalk@msu.edu}
Darryl Z. Seligman,$^{1}$
\\
$^{1}$Department of Physics and Astronomy, Michigan State University, East Lansing, MI 48824, USA\\
}

\pubyear{2026}

\begin{document}
\label{firstpage}
\pagerange{\pageref{firstpage}--\pageref{lastpage}}
\maketitle

\begin{abstract}
In the past decade, three macroscopic-scale interstellar objects have been discovered, implying that a larger galactic population exists. In this paper, we investigate the possibility that the rapid sky-plane velocities of interstellar objects may preclude their discovery. We provide an analytic solution for the apparent sky motion of an object on an arbitrary orbit observed at an arbitrary location which is more efficient and requires less overhead than the numerical approach. This formula is applied to evaluate the typical sky motion of an interstellar object as a function of its orbit and limiting magnitude/distance. We generate three synthetic populations of $\sim10^5$ interstellar objects within heliocentric spheres of radii 1.2, 3.0, and 5.0 AU, and calculate the sky motion for these objects when they reach a range of limiting magnitudes for multiple populations of interstellar asteroids and comets. The sky motions of the three known interstellar objects are broadly characteristic of populations with similar absolute magnitudes. Moreover, the intrinsically brighter objects reach detection magnitude thresholds at lower on-sky speeds than the dim objects, and active comets at even lower speeds for the same apparent magnitudes. The tails of these distributions extend to speeds faster than the discovery motion of 1I. Therefore, the potentially rapid sky motion of interstellar objects should be taken into account when attempting to link hyperbolic trajectories in survey data.

\end{abstract}


\section{Introduction}
To date, there have been three macroscopic, $\sim$100m to $\sim$km scale interstellar objects discovered traversing the inner Solar System on hyperbolic trajectories. It has been established, primarily based on the existence and structure of the Kuiper belt and Oort cloud, that the formation and subsequent dynamical evolution of the Solar System ejected a population of interstellar comets \citep{Hahn1999,Gomes2004,Tsiganis2005,Morbidelli2005,Nesvorny2018}. {A natural extrapolation is that extrasolar planetary systems also produce interstellar comets.} {These objects were theorized to exist and to have passed through our Solar System undetected for many decades prior to their discovery \citep{sekanina1976probability,mcglynn1989,Sen1993,Jewitt2003,Engelhardt2014,Cook2016,Moro2009}. }

Despite their theorized existence, no interstellar objects were detected until the discovery of 1I/`Oumuamua in 2017 \citep{Williams17} by the PanSTARRS survey \citep{Chambers2016}. Since the discovery of the first interstellar object, only two more have been found: 2I/Borisov in 2019 \citep{borisov_2I_cbet} and 3I/ATLAS in 2025 \citep{Denneau2025} by ATLAS \citep{Tonry2018a,Tonry2018b}.

The three interstellar objects discovered to date displayed divergent properties which nontrivially contributed to the circumstances leading to their discovery. 1I/`Oumuamua defied many expectations; it was photometrically inactive and displayed no coma or visible cometary tail \citep{Meech2017,Ye2017,Jewitt2017,Trilling2018}. {Therefore, this object was intrinsically dim and only detected (and detectable) when it was extremely close to the Earth, under optimal and extremely fortuitous discovery circumstances.} 1I exhibited significant nongravitational acceleration \citep{Micheli2018}. It displayed brightness fluctuations of $\sim2.5-3.5$ magnitudes corresponding to a $6:6:1$ geometry \citep{Meech2017,Knight2017,Bannister2017,Fraser2017,Belton2018,Mashchenko2019,Taylor2023} and a moderately red color \citep{Meech2017,Fitzsimmons2017,Ye2017}. It has been suggested that the extreme fluctuations in brightness contributed to its overall detectability, since 1I was detected at a peak in intrinsic brightness (Figure 1 of \citet{Belton2018}); however, the effects of apparent magnitude fluctuations of objects with extreme shapes have been shown to statistically bias these populations against discovery \citep{Levine2023}. The combination of comet-like nongravitational acceleration and lack of visible tail or coma led to a variety of hypotheses regarding its origins, invoking radiation pressure \citep{MoroMartin2019,Bialy2018,Flekkoy19,Luu20} or outgassing with little dust \citep{Micheli2018,Sekanina2019,Seligman2020,Levine2021,Levine2021_h2,desch20211i,jackson20211i,Desch2022,Bergner2023}. Since then, a population of photometrically inactive small bodies in the near-Earth environment have been identified to also have comet-like nongravitational accelerations \citep{Farnocchia2023,Seligman2023b,Seligman2024PNAS}. This new population of `dark comets' implies that objects like 1I are more common than previously thought.

Both 2I/Borisov in 2019 \citep{Jewitt2019b,Fitzsimmons2019,Ye2019,McKay2020,Guzik2020,Hui2020,Kim2020,Cremonese2020,yang2021} and 3I/ATLAS in 2025 \citep{Seligman2025,Opitom2025,Kareta2025,Frincke2025,Jewitt2025,Alarcon2025,Belyakov2025,Hoogendam2025,Jewitt2025_NOT,Tonry2025,Marcos2025} were discovered contemporaneously with activity, as is typical of cometary bodies. Despite having $\sim$km scale nuclei, the brightening enhancement from cometary activity rendered both objects detectable at significantly further heliocentric (and geocentric) distances than 1I/`Oumuamua. Spectroscopic observations of these objects demonstrated that both have a composition with some discernible differences to that of typical solar system comets, highlighted by their hypervolatile-enriched compositons \citep{Cordiner2020,Bodewits2020,Cordiner2025,Lisse2025,Yang2025,Xing2025,SalazarManzano2025,Paek2026,Coulson2026,Lazzarin2026,Roth2025,opitum2026,manzano2026,roth2026,Cordiner_2026}. The abundance of hypervolatiles presumably contributed to their activity and subsequent detectability at larger distances. For reviews on interstellar objects discovered passing through the Solar System prior to 3I/ATLAS, we refer the reader to \citet{MoroMartin2022,Jewitt2023ARAA,Seligman2023,Fitzsimmons2024}.

The kinematics of the three known interstellar objects are also different. The excess velocity of 1I/`Oumuamua, 2I/Borisov, and 3I/ATLAS were $V_\infty\sim26$ km s$^{-1}$, $V_\infty\sim32$ km s$^{-1}$, and $V_\infty\sim58$ km s$^{-1}$ respectively. These approximately correspond to ages of $\sim10^2$, $\sim10^3$, and $\sim10^4$~Myr, albeit with large uncertainty \citep{Mamajek2017,Gaidos2017a, Feng2018,Fernandes2018,Hallatt2020,Hsieh2021,Hopkins2025,Taylor2025}. Naively, the combination of this rapid orbital motion and possible retrograde geometry makes interstellar objects move more rapidly across the sky than typical solar system objects. While \citet{MarcetaSeligman2023} calculated the sky-plane velocity of a subset of synthetic interstellar objects that pass a detectability criteria (Figure 2 in that work), there has been no calculation to date of the typical interstellar object rate of sky motion. This rapid sky motion, and the associated trailing loss, may render typical interstellar objects undetectable. While we do not account for the effects of trailing losses in this work, this paper lays the foundation for future works to quantify the capability of observatories and all-sky surveys in discovering new interstellar objects.
    
The detection of 1I implies there should $\sim1$ of these objects traversing the inner Solar System at any given time \citep{Laughlin2017,Jewitt2017,Do2018,MoroMartin2018i,Moro2019exOC,Trilling2017}. Moreover, the large nuclear radius of 3I/ATLAS implies an ever greater number density, and that interstellar objects must have been missed by surveys \citep{Hui2026}.

There have been several precovery observations of the known interstellar objects that were either attempted and/or successful. A precovery search of 1I/`Oumuamua in data from the Solar and Heliospheric Observatory (SOHO) and Solar TErrestrial RElations Observatory (STEREO) provided nondetections \citep{Hui2019}. However, precovery observations of 2I/Borisov were successfully reported in the Zwicky Transient Facility (ZTF) \citep{Ye2020}. 3I/ATLAS was also apparent in several facilities before its discovery, specifically in the Rubin Observatory Legacy Survey of Space and Time (LSST) images \citep{Chandler2025}, NASA's Transiting Exoplanet Survey Satellite (TESS) images \citep{Feinstein2025,Martinez-Palomera2025}, and ZTF \citep{Ye2025}. This is indicative that more of these objects may persist in data unnoticed. 

Therefore, it is possible that we are missing interstellar objects in our surveys. However, in this manuscript we intentionally do not speculate on what aspects of discovery pipelines may inhibit interstellar object discoveries. In this paper, we investigate whether large rates of sky motion could be partially responsible for this, given difficulties associated with linking rapidly moving objects. To that end, we present an analytic formula for the apparent sky motion of an interstellar object as a function of its orbital elements, also accounting for the position of the Earth. As opposed to direct numerical integration, this approach provides a rapid method for approximating the plane-of-sky motion of a solar system body as a function of its orbit and Earth's position. This method is more efficient than direct numerical integration, does not require the overhead of setting up numerical integrations and a framework for calculating on-sky positions, and is generic to any moving object observed from any location. We then create a synthetic population of $\sim10^5$ interstellar objects using the probabilistic method \citep{Marceta2023} and calculate the distribution of their apparent rates of sky motion at a range of limiting  magnitudes for inactive objects and active comets. We have not rigorously simulated survey detection limits in this work as performed, for example, by \citet{MarcetaSeligman2023,Hoover2022,Dorsey2025}, but instead placed cutoffs based on how bright these objects appear as viewed from Earth. This is because we aim to quantify the distribution of sky motions for objects on hyperbolic orbits for a generic survey, and independent of the specifics of a given survey. Although we do not include survey-specific effects in this analysis, this analytic method can help all-sky surveys understand their capabilities in regard to interstellar object sky motions.

\section{Analytic Formula for Sky Motion} \label{sec: Analytic_Formula}
In this section, we formulate an equation to calculate the apparent rate that an object on a generic orbit moves across the observing body's sky. We apply this equation specifically to the case of a hyperbolic orbit for an interstellar object as viewed from Earth. 
\subsection{Generalized Vector Form} \label{subsec: gen_vector_form}
We aim to formulate an analytic expression for the apparent angular rate of motion of an interstellar object across the plane of sky as viewed from Earth. We begin by generalizing our calculations in vector form, and then extend to 2D and 3D cases as functions of the object's Keplerian orbital elements. 
If we subtract the component of the interstellar object's orbital velocity vector, $\vec{{v}}$ along a line of sight from Earth, we can find the transverse velocity at which the object moves across our sky, $\vec{{v}}{_{sky}}$, using:
\begin{equation} 
    \vec{{v}}{_{sky}} = \vec{{v}} - (\vec{{v}} \cdot \hat{{d}})\hat{{d}} \, ,
    \label{eqtn: onsky_linearvelocity}
\end{equation} 
{where $\hat{d}$ is the line-of-sight direction and the unit-normalized vector of $\vec{{d}}$, the separation vector between the Earth and the object, }

\begin{equation} 
    \vec{{d}}=\vec{r} - \vec{r}_\oplus \, .
    \label{eqtn: d_separation}
\end{equation}

The vectors $\vec{r}$ and $\vec{r}_\oplus$ are the heliocentric distance vectors of the object and Earth, respectively.
From this, we can form a vector equation for the angular rate of sky motion using this distance and assuming a small angle $\theta$,
\begin{equation} 
    \frac{\Delta \theta}{\Delta t} = \frac{|\vec{{v}}{_{sky}}|}{|\vec{{d}}|} \, .
    \label{eqtn: skymotion_setup}
\end{equation} 
Substituting Equation \ref{eqtn: onsky_linearvelocity} into Equation \ref{eqtn: skymotion_setup}, we obtain our vector equation for angular rate of sky motion, 
\begin{equation} 
    \frac{d\theta}{dt} = \bigg|\frac{\vec{{v}}}{|\vec{{d}}|} - \frac{(\vec{{v}} \cdot \vec{{d}})\vec{{d}}}{|\vec{{d}}|^3} \bigg| \, .
    \label{eqtn: skymotion_vectoreqtn}
\end{equation}

For the purpose of this paper, in Equation 4, we take the absolute value of the $d\vec{\theta}/dt$ vector, as we aim to quantify the magnitudes of the rates at which interstellar objects should appear to move across the sky, $d\theta/dt$. Future analyses that include survey-specific effects that depend on the direction of interstellar objects moving across the sky should use the full vector form of $d\vec{\theta}/dt$, represented by the vector quantity inside the absolute value bars in Equation 4.

\subsection{2D Function of Orbital Elements} \label{subsec: 2D_skymotion}
We begin by assuming a general hyperbolic orbit with zero inclination. This makes our problem strictly 2D.
The goal is to calculate this rate of angular sky motion using the 2D orbital elements of the object: semimajor axis $a$, eccentricity $e$, and true anomaly $f$ - as well as the position of the Earth, denoted strictly from its true anomaly value $f_\oplus$, assuming a circular orbit with a 1 AU radius. 
This equation will therefore apply to any object orbiting in the Earth-Sun plane, though the hyperbolic orbit of an interstellar object has an unphysical subset of true anomalies, which we take into account in later applications of our formula.

We start by writing the distance vector $\vec{d}$ as a function of orbital elements by substituting the Keplerian heliocentric distance for the interstellar object into Equation \ref{eqtn: d_separation}. However, we want to generalize over any relative arguments of periastron, so we use the perifocal 2D cartesian reference frame with the $x$-axis aligned the interstellar object argument of periastron $\omega$. Since we assume the Earth's orbit is circular, its argument of periastron is arbitrary; thus, we can assume the object and Earth share an argument of periastron $\omega_\oplus=\omega=0$ with respect to our defined $x$-axis. The distance vector from Earth to an interstellar object in this reference frame is:
\begin{equation} 
    \begin{split}
    \vec{{d}}=  \Bigg[\bigg|\frac{a(1-e^2)}{1+e\cos f}\bigg|\cos f - r_\oplus\cos f_{\oplus}\Bigg]\hat{x}\,\, \\+ \Bigg[\bigg|\frac{a(1-e^2)}{1+e\cos f}\bigg| \sin f - r_\oplus\sin f_{\oplus}\Bigg]\hat{y}\, ,  
    \end{split}
    \label{eqtn: d_xy_elements}
\end{equation}
where we assume the heliocentric distance of Earth is $r_\oplus=a_\oplus=1$ AU, and the heliocentric distance of the interstellar object is: 
\begin{equation}
    r = \bigg|\frac{a(1-e^2)}{1+e\cos f}\bigg| \, ,
    \label{eqtn: r_obj_elements}
\end{equation}
where $a$, and thus $r$, have units of AU. \\
We then solve for the orbital velocity vector of the interstellar object in terms of its semimajor axis $a$, eccentricity $e$, and true anomaly $f$, using orbital angular momentum conservation, which also applies to the hyperbolic case, although the object will have some excess velocity $v_\infty$. 
The specific angular momentum, $h$, of an orbit is:
\begin{equation}
    h = \sqrt{GM|a(1-e^2)|}\, .
    \label{eqtn: h_specific_angmom}
\end{equation}
In Equation \ref {eqtn: h_specific_angmom}, $G$ is the gravitational constant and $M$ is the mass of the Sun. Therefore, our expression for the orbital velocity vector as a function of orbital elements in our cartesian coordinate basis in the object's perifocal reference frame becomes:
\begin{equation}
    \vec{{v}} = \sqrt\frac{GM}{|a(1-e^2)|}\Big[-\sin f \hat{x} + (e+\cos f)\hat{y}\Big] \, .
    \label{eqtn: v_xy_elements}
\end{equation}
We can now substitute Equations \ref{eqtn: d_xy_elements} and \ref{eqtn: v_xy_elements} into our generalized vector form from Equation \ref{eqtn: skymotion_vectoreqtn} to obtain a 2D analytic formula for the angular rate of sky motion as a function of orbital elements,
\begin{equation}
\begin{split}
    \frac{d\theta}{dt} = \sqrt\frac{GM}{|a(1-e^2)|}\bigg|\frac{r(1+e\cos f)-r_\oplus(\cos(f-f_\oplus) +e\cos f_\oplus)}{r^2-2rr_\oplus\cos(f-f_\oplus)+r_\oplus^2}\bigg|\, , 
    \end{split}
    \label{eqtn: skymotion_orbparams}
\end{equation}
written, for simplicity, using the heliocentric distance of the interstellar object, $r$, from Equation \ref{eqtn: r_obj_elements}, and assuming $r_\oplus=a_\oplus=1$ AU for a circular Earth orbit with $f_\oplus$ {defined} with respect to the $x$-axis. This formula can, in general, be applied to both bound and unbound orbits (excluding the specific parabolic case, $e=1$), as we account for the hyperbolic case ($e>1$) using the absolute value of the semi-latus rectum, $p=a(1-e^2)$. However, in the hyperbolic regime, there would be inherent bounds on the range of possible true anomalies for these orbits, defined by:
\begin{equation}
|f|_{r\rightarrow\infty} = \arccos\bigg(-\frac{1}{e}\bigg) \, . 
\label{eqtn: truanom_limit}
\end{equation}

\begin{figure}
    \centering
    \includegraphics[width=\linewidth]{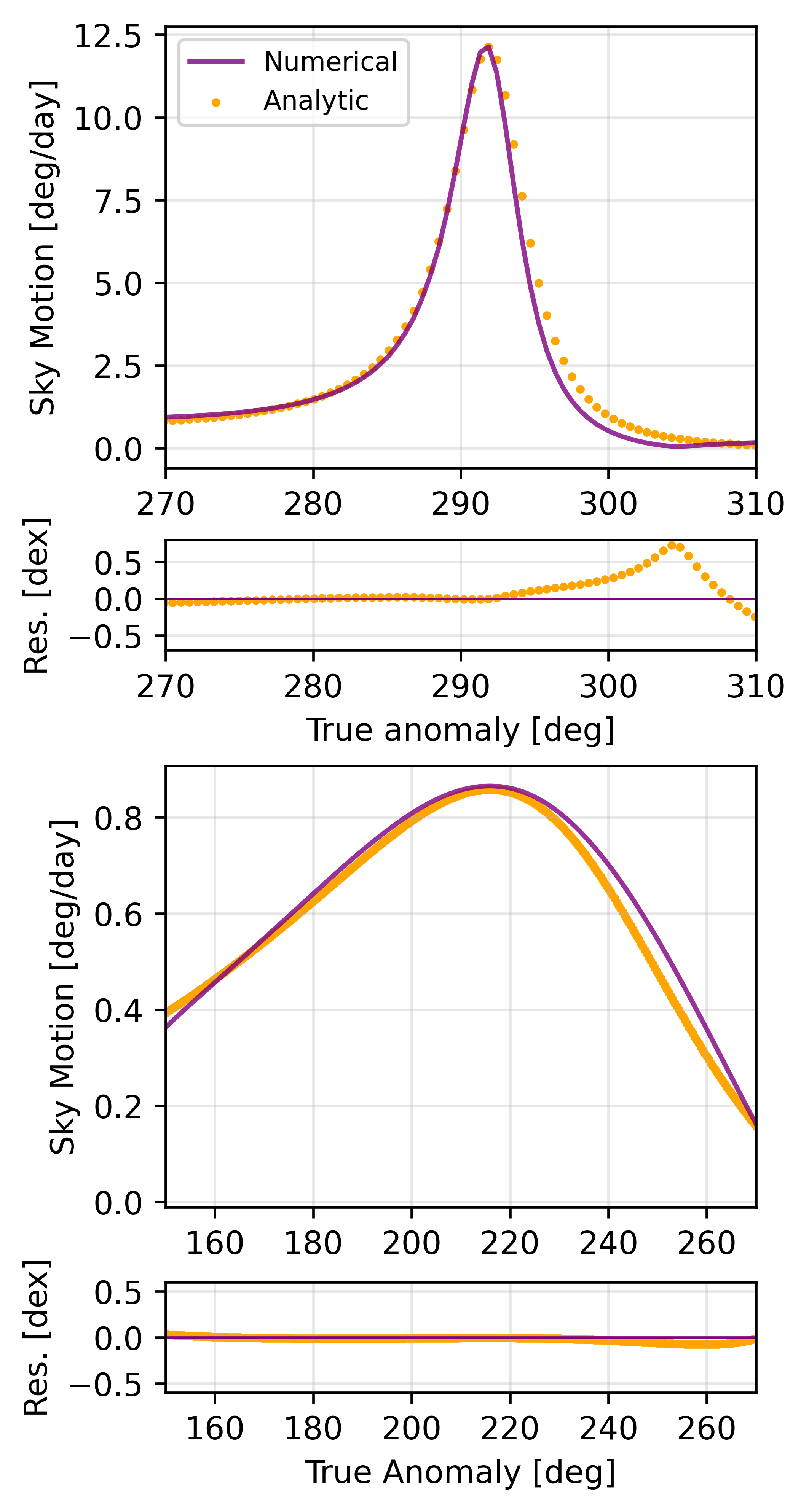}
    \caption{Sky motion through the trajectories of 1I/`Oumuamua (top) and 2I/Borisov (bottom),
    using JPL Horizons ephemerides data, overplotted with values from 3D calculation in Equation \ref{eqtn:3D_skymotion_orbparams_full}.}
    \label{fig:knownISOmotion_1Dplot}
\end{figure}

\begin{table}
\begin{center}
\begin{tabular}{ccccc}
\hline
\multicolumn{1}{c}{Object} & \multicolumn{1}{c}{Discovery Date} & \multicolumn{1}{c}{$m_V$} & \multicolumn{1}{c}{Sky Motion [$^\circ$/d]} & \multicolumn{1}{c}{$r$ [AU]}\\
\hline
1I & 19 Oct 2017 & 19.705 & 6.64 & 1.21 \\
2I & 29 Aug 2019 & 18.815 & 0.47 & 3.00 \\
3I & 01 Jul 2025 & 18.170 & 0.48 & 4.51 \\
\hline
\multicolumn{1}{c}{Object} & \multicolumn{1}{c}{Fastest Date} & \multicolumn{1}{c}{$m_V$} & \multicolumn{1}{c}{Max. Motion [$^\circ$/d]} & \multicolumn{1}{c}{$r$ [AU]}\\
\hline
1I & 15 Oct 2017 & 19.958 & 12.21 & 1.11 \\
2I & 08 Dec 2019 & 16.663 & 0.86 & 2.01 \\
3I & 26 Dec 2025 & 15.578 & 1.24 & 2.49 \\
\hline
\end{tabular}
\caption{Apparent magnitude, sky motion, and heliocentric distance of the three known interstellar objects: 1I/`Oumuamua, 2I/Borisov, and 3I/ATLAS at the time of discovery and the time of their fastest sky motion. All values are gathered from the JPL Horizons database.}
\label{table: known_ISO_values_discovery_max} 
\end{center}
\end{table}

\begin{figure}
    \centering
    \includegraphics[width=\linewidth]{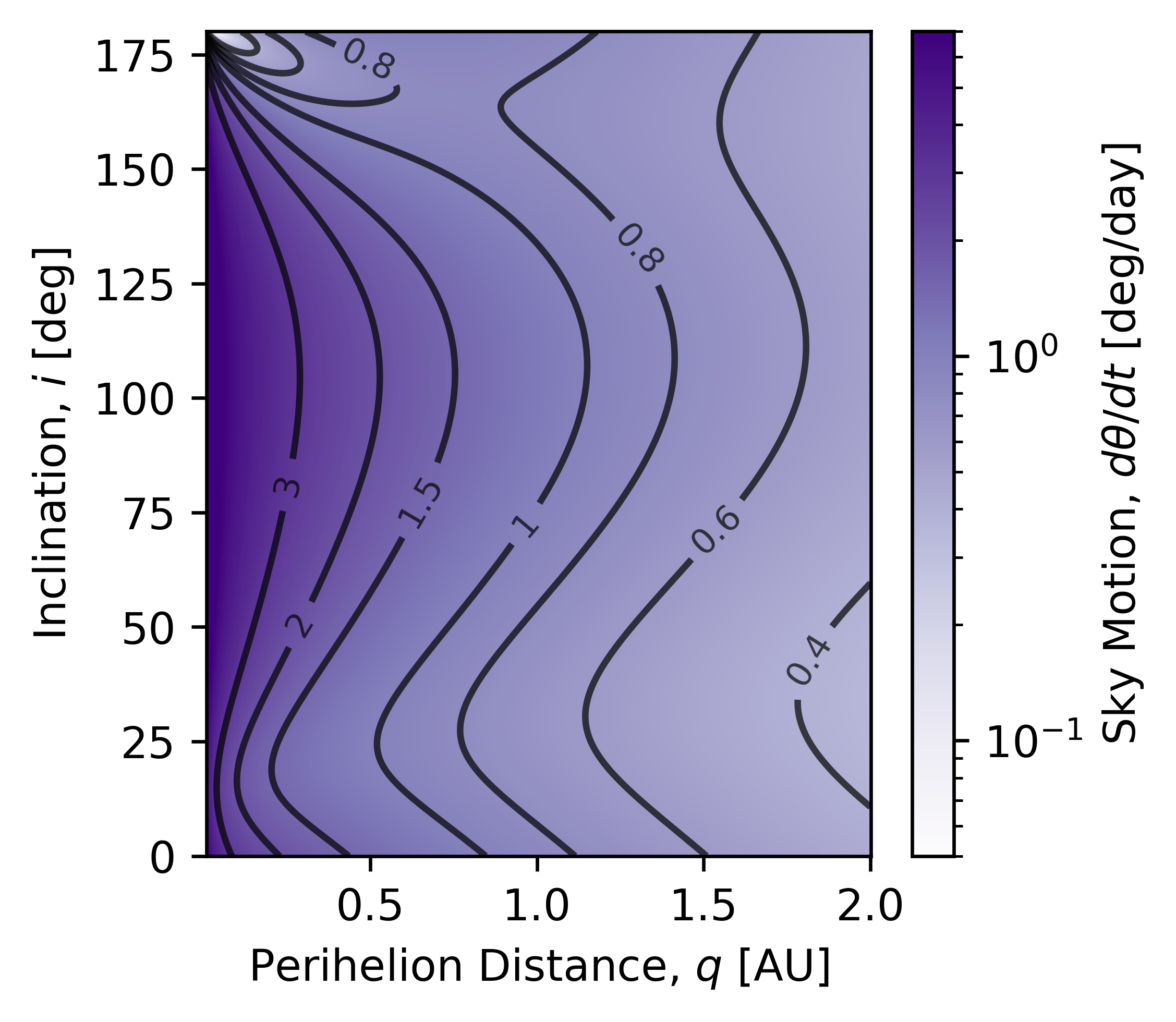}
    \caption{The apparent sky motion as a function of perihelion distance and orbital inclination. We set remaining parameters of interstellar object eccentricity, argument of periastron, and longitude of ascending node, to be $e = 1.9$, $\omega = 3.4$ rad, and $\Omega=3.1$ rad based on the spread of values for these elements in the synthetic population generated in Section \ref{subsec: Synthetic_Population}. The true anomaly values of the interstellar object and Earth are taken to be $f = f_\oplus = 0.0$ rad, corresponding to perihelion, though by using the true values of Earth's orbital elements, these are not necessarily aligned.}
    \label{fig:sky_motion_qvi}
\end{figure}

\begin{figure}
    \centering
    \includegraphics[width=\linewidth]{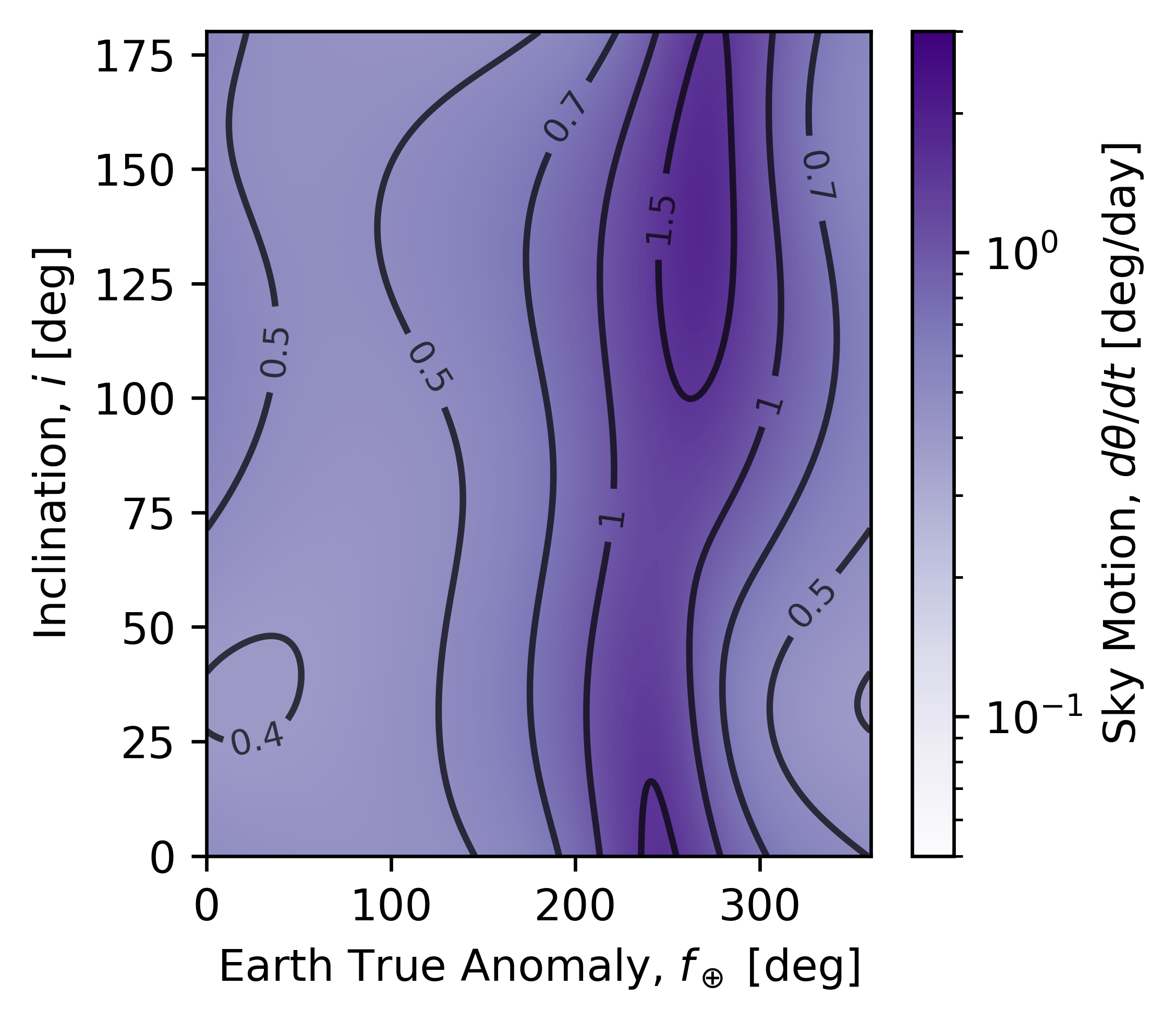}
    \caption{Apparent sky motion for given values of the true anomaly of the Earth and the interstellar object orbital inclination. We set the remaining parameters of interstellar object semimajor axis, eccentricity, true anomaly, argument of periastron, and longitude of ascending node, to be $a=-2.0$ AU, $e = 1.9$, $f=0.0$ rad, $\omega = 3.4$ rad, and $\Omega=3.1$ rad based on the spread of values for these elements in the synthetic population generated in Section \ref{subsec: Synthetic_Population}.}
    \label{fig:sky_motion_fevi}
\end{figure}

\subsection{3D Function of Orbital Elements} \label{subsec: 3D_skymotion}
Next, we expand our previous equation to apply to a general 3D orbit. We start by defining the position of the interstellar object in its own perifocal cartesian reference frame,
\begin{equation}
\vec{r} = r[\cos f\hat{x} + \sin f\hat{y} + 0\hat{z}] \, ,
\label{eqtn: ISO_position_perifocal}
\end{equation}
where the $x$-axis is defined to be along the argument of perihelion in the plane of the object's orbit, and $r$ is the heliocentric distance of the object, as in Equation \ref{eqtn: r_obj_elements}.
In the same reference frame, the velocity of the object is given by Equation \ref{eqtn: v_xy_elements}.
We then aim to transform these expressions into a new coordinate basis {$\hat{x}^\prime,\hat{y}^\prime,\hat{z}^\prime$} in the Sun's ecliptic plane, in which the $x^\prime$-axis is defined along the vernal equinox, or first point of Aries, and the $z^\prime$-axis is defined to be perpendicular to the ecliptic plane. This transformation requires three rotations of the original reference frame. 
First, we rotate the ${x,y,z}$ coordinate system about the $z$-axis by $-\omega$, to align the $x$-axis, previously along the object's argument of perihelion, with the longitude of the ascending node $\Omega$ in the ecliptic plane with the rotation matrix defined by:
\begin{equation}
\overleftrightarrow{{Rot}}_1(-\omega) =
\begin{pmatrix}
\cos(-\omega) & -\sin(-\omega) & 0\\
\sin(-\omega) & \cos(-\omega) & 0\\
0 & 0 & 1
\end{pmatrix} \ .
\label{eqtn: Rot1_matrix_w}
\end{equation}
Next, we rotate about the $x$-axis by $-i$, to transform the object's inclined orbit into the ecliptic plane with the rotation matrix defined by:
\begin{equation}
\overleftrightarrow{{Rot}}_2(-i) =
\begin{pmatrix}
1 & 0 & 0\\
0 & \cos(-i) & -\sin(-i)\\
0 & \sin(-i) & \cos(-i)
\end{pmatrix} \ .
\label{eqtn: Rot2_matrix_i}
\end{equation}
Finally, we rotate about the $z$-axis by $-\Omega$ to align the object's ascending node with the vernal equinox, as the longitude of the ascending node is typically defined with respect to the vernal equinox, with the rotation matrix defined by:
\begin{equation}
\overleftrightarrow{{Rot}}_3(-\Omega) =
\begin{pmatrix}
\cos(-\Omega) & -\sin(-\Omega) & 0\\
\sin(-\Omega) & \cos(-\Omega) & 0\\
0 & 0 & 1
\end{pmatrix} \ .
\label{eqtn: Rot3_matrix_Om}
\end{equation}
Thus, the total rotation matrix is:
\begin{equation}
\overleftrightarrow{{M}} = \overleftrightarrow{{Rot}}_3(-\Omega)\overleftrightarrow{{Rot}}_2(-i)\overleftrightarrow{{Rot}}_1(-\omega) \, ,
\label{eqtn: Full_Rotation_Matrix_unsimplified}
\end{equation}
and the fully simplified version of this matrix can be found in the appendix (Equation \ref{eqtn: Full_Rotation_Matrix}).

We apply these rotation matrices to both the position and velocity vectors {(Equations \ref{eqtn: ISO_position_perifocal} and \ref{eqtn: v_xy_elements}, respectively)} and recover the expressions for these quantities in our ecliptic plane coordinate basis as: 
\begin{equation}
\begin{split}
\vec{r^\prime} = \bigg|\frac{a(1-e^2)|}{1+e\cos f}\bigg|\bigg[\Big[\sin{f}(\cos{\Omega}\sin{\omega}+\cos{i}\cos{\omega}\sin{\Omega})\,\,\,\,\,\,\,\,\,\,\,\, \\
    + \cos{f}(\cos{\Omega}\cos{\omega}-\cos{i}\sin{\omega}\sin{\Omega})\Big]\hat{x}^\prime \,\,\,\,\,\,\\
    + \Big[\cos{i}\cos{\Omega}\sin({f-\omega}) - \sin{\Omega}\cos({f-\omega})\Big]\hat{y}^\prime \,\,\,\,\,\,\,\\
    + \Big[\sin{i}\sin({f-\omega})\Big]\hat{z}^\prime\bigg]\,,
\end{split}
\label{eqtn: rvec_prime3D}
\end{equation}

and

\begin{equation}
\begin{split}
\vec{{v}^\prime} =
\sqrt\frac{GM}{|a(1-e^2)|}\bigg[\Big[\cos{\omega}((e+\cos{f})\cos{i}\sin{\Omega}-\cos{\Omega}\sin{f}) \,\,\,\,\,\,\,\,\,\\ 
+ \sin{\omega
}((e+\cos{f})\cos{\Omega}+\sin{f}\cos{i}\sin{\Omega})\Big]\hat{x}^\prime\,\,\,\,\\
+ \Big[\cos{i}(\cos{(f-\omega)}+e\cos{\omega})\cos{\Omega}\,\,\,\,\,\,\,\,\,\, \\
+ (\sin{(f-\omega)}-e\sin{\omega})\sin{\Omega}\Big]\hat{y}^\prime \,\,\,\,\,\,\\
+ \Big[-\sin{i}(\cos({f-\omega})+e\cos{\omega}\Big]\hat{z}^\prime\bigg] \, .
\end{split}
\label{eqtn: vvec_prime3D}
\end{equation}

In this coordinate basis, Earth's heliocentric distance vector is,
\begin{equation}
\vec{r^\prime}_\oplus = \frac{a_\oplus(1-e_\oplus^2)}{1+e_\oplus\cos f_\oplus}\Big[\cos{(f_\oplus + \omega_\oplus)}\hat{x}^\prime + \sin{(f_\oplus + \omega_\oplus)}\hat{y}^\prime + 0\hat{z}^\prime\Big] \, ,
\label{eqtn: rearthvec_prime3D}
\end{equation}
and we use values of Earth's semimajor axis, eccentricity, and argument of periastron with respect to the vernal equinox: $a_\oplus=1.000003$ AU, $e_\oplus=0.01671$, and $\omega_\oplus=102.9^\circ=1.795$ rad. 

We plug Equations \ref{eqtn: rvec_prime3D} and \ref{eqtn: rearthvec_prime3D} into Equation \ref{eqtn: d_separation} to find the distance vector from Earth to an interstellar object. Then, we plug this calculated distance vector and Equation \ref{eqtn: vvec_prime3D} into Equation \ref{eqtn: skymotion_vectoreqtn} to find the angular rate of sky motion for a general 3D orbit. The full equation has a closed form solution, and is included in the appendix (Equation \ref{eqtn:3D_skymotion_orbparams_full}). It should be noted that this equation is undefined when the distance between the Earth and the object is $d=0$, and also in the hyperbolic regime ($e>1$, $a<0$) when $f>\arccos(-1/e)$, or beyond the true anomaly limit defined in Equation \ref{eqtn: truanom_limit}.

To further validate our calculations, we apply our final 3D equation to the orbits of the first two discovered interstellar objects. We plot these comparisons against orbital ephemeris data from JPL Horizons in Figure \ref{fig:knownISOmotion_1Dplot}, and demonstrate that our results map very closely to the measured data, with residuals largely on the order of $\sim10^{-2}$. As the object gets closer to the Earth, its apparent sky motion increases and then decreases as it continues to get farther. The divergence of our calculation from the JPL data is due to the fact that we assume the Earth is stationary as the interstellar object travels through its orbit. Nongravitational accelerations experienced by the object, which are not modeled in our analytic formula, could also contribute to the divergence from the numerical solution, though these would have relatively minor effects. The dependence on Earth position is much higher for 1I, as it passed very close to Earth ($d\sim0.16$ AU), causing any small change in the Earth's orbit to be exacerbated, resulting in relatively higher residuals. We recognize that the position of Earth in the orbits would introduce significantly larger differences than the difference between the numerical and analytic prescriptions; however, our population-level analysis marginalizes over Earth positions. The sky motions and apparent magnitudes of the three known interstellar objects at the date of their discovery, as well as their fastest sky motions, collected from the JPL Horizons database, are compiled in Table \ref{table: known_ISO_values_discovery_max}.

We present some of the structure of Equation \ref{eqtn:3D_skymotion_orbparams_full} in Figures \ref{fig:sky_motion_qvi} and \ref{fig:sky_motion_fevi}. Figure \ref{fig:sky_motion_qvi} demonstrates that farther objects are indeed moving slower across the sky, and that a 180 degree inclination minimizes the sky motion in the ecliptic plane by misaligning the perihelion vectors of the Earth and the object. Note that in Figure 2, the values of sky motion are not the same at $i=0^\circ$ and $i=180^\circ$, due to use of nontrivial values for the object orbital elements of $a$, $e$, $\omega$, and $\Omega$, as well as the true values for the orbital elements of Earth. Figure \ref{fig:sky_motion_fevi} demonstrates how sky motion increases when the Earth and interstellar object reach points of conjunction, regardless of the inclination of the object.

\section{Application \& Results} \label{sec: Applications_Results}
\subsection{Synthetic Interstellar Object Population}\label{subsec: Synthetic_Population}

\begin{figure*}
    \centering
    \includegraphics[width=\linewidth]{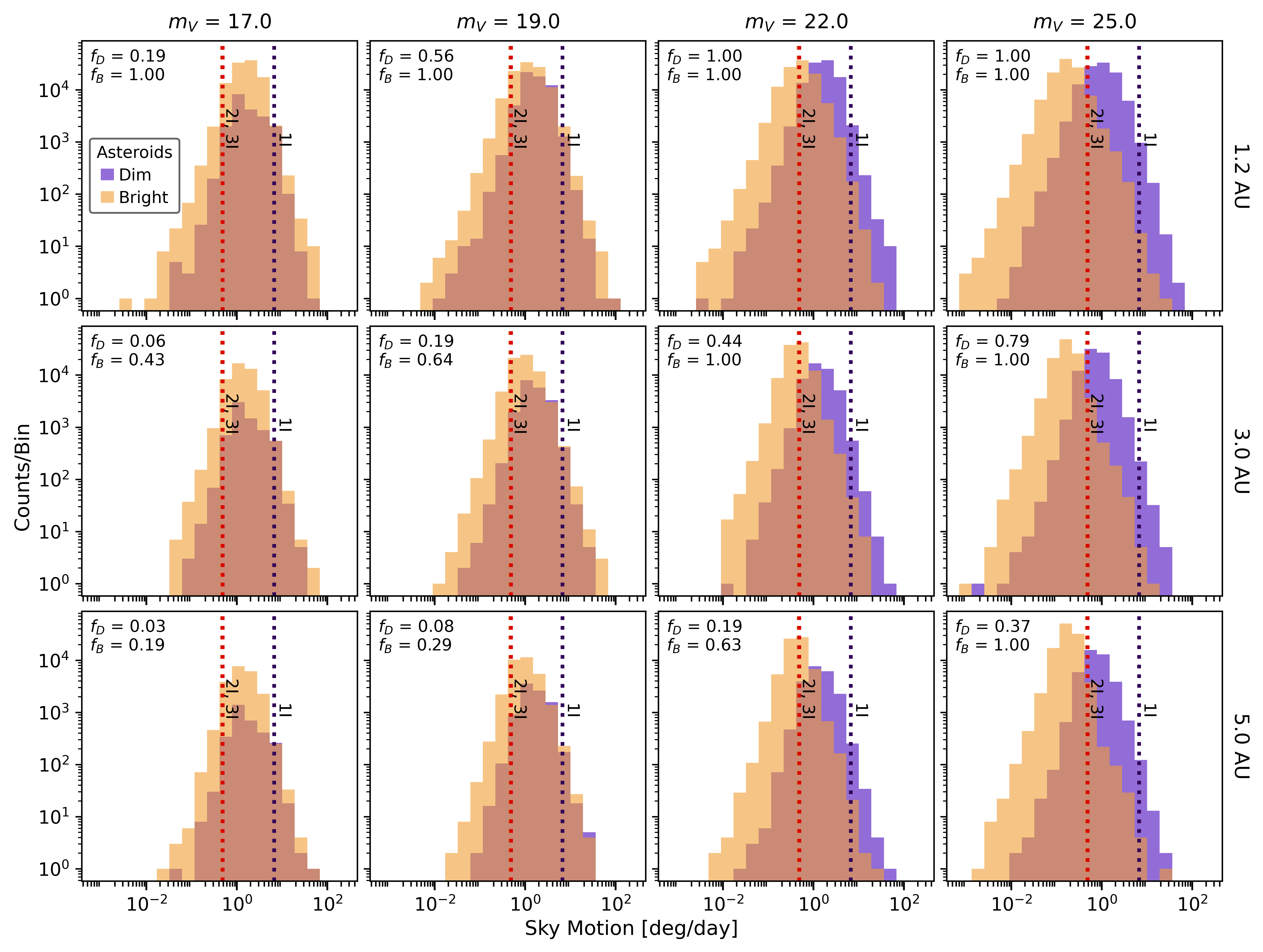}
    \caption{{(\textit{Top row}) Distributions of sky motions for asteroid-like objects, calculated using the 3D approximation in Equation \ref{eqtn:3D_skymotion_orbparams_full} for 106,398 interstellar objects within a heliocentric model sphere of radius 1.2 AU when they reach apparent v-band magnitudes of 17.0, 19.0, 22.0, and 25.0. The red dotted lines mark the apparent sky motion of 1I/`Oumuamua at discovery (6.64 deg/day), and the black dotted lines represent the apparent sky motions of 2I/Borisov and 3I/ATLAS at discovery (0.47 and 0.48 deg/day, respectively). The purple/yellow distributions represent the populations of objects whose apparent magnitudes were calculated using $H_V=22.08$/$H_V=17.1$, the absolute magnitude of 1I/3I. The fraction of these intrinsically dim/bright objects that reach each given magnitude is denoted on the panels by $f_D$/$f_B$. (\textit{Middle row}) Same as top row, but for 104,225 interstellar objects within a 3.0 AU heliocentric sphere. (\textit{Bottom row}) Same as top and middle rows, but for 106,998 interstellar objects within a 5.0 AU heliocentric sphere.}}
    \label{fig:ALL_limit_motions}
\end{figure*}

To apply our calculation to a statistically robust number of interstellar objects requires a synthetic population, as we have only detected three interstellar objects at the time of writing. {We generate our population using the probabilistic method \citep{Marceta2023}, which generates a number of synthetic interstellar objects in a given heliocentric model sphere, based on input stellar kinematic parameters.} For the purposes of this analysis, we use M-star kinematics. The kinematic distribution of typical interstellar objects is generally unconstrained, although theoretical investigations have considered various kinematics \citep{Lintott2022,Hopkins2023,Hopkins2025b,Forbes2025}. We generate three populations of $\sim10^5$ interstellar objects, each within a different heliocentric model sphere with radii of 1.2 AU, 3.0 AU, and 5.0 AU, representing (approximately) near-Earth objects, inner solar system objects, and Jupiter-distance objects, respectively.

\subsection{Interstellar Asteroid Populations} \label{subsec: Analysis}

For each interstellar object in each of our three populations (1.2, 3.0, and 5.0 AU), we calculate the apparent magnitude of the object at 100 positions through the object's hyperbolic trajectory, defined by a range of true anomaly values as posited in Equation \ref{eqtn: truanom_limit},
\begin{equation}
    f \in \Bigg(-\arccos\bigg(-\frac{1}{e}\bigg), +\arccos\bigg(-\frac{1}{e}\bigg)\Bigg) \, .
    \label{eqtn: trueanom_hyprange}
\end{equation}

The apparent magnitude at each position is calculated using the method in \cite{Seligman2018} and \cite{Hoover2022}:
\begin{equation}
    m = H + 2.5 \textrm{log}_{10}\bigg(\frac{{d}_{OS}^2{d}_{OE}^2}{q(\alpha){d}_{ES}^4}\bigg) \, ,
    \label{eqtn: AppMag_SS}
\end{equation}
where $H$ is the absolute magnitude of the object, and the parameters $d_{OS}$, $d_{OE}$, and $d_{ES}$ are the Object-Sun distance, Object-Earth distance, and Earth-Sun distance, respectively. The phase angle $\alpha$ and phase integral $q(\alpha)$ used to calculate apparent magnitudes of bodies in the Solar System are defined by:
\begin{equation}
    \cos \alpha = \frac{{d}_{OS}^2 + {d}_{OE}^2 - {d}_{ES}^2}{2{d}_{OS}{d}_{OE}} \, ,
    \label{eqtn: cosine_phaseangle}
\end{equation}
{and}
\begin{equation}
    q(\alpha) = \frac{2}{3}\Bigg(\bigg(1-\frac{\alpha}{\pi}\bigg)\cos\alpha+\frac{1}{\pi}\sin\alpha\Bigg) \, .
    \label{eqtn: phase_function}
\end{equation}
The apparent magnitude calculation requires an assumption of each object's absolute magnitude H. For this calculation, we test two values of absolute magnitude, corresponding to an intrinsically dim population and an intrinsically bright population. For the dim population, we use the absolute magnitude of 1I/`Oumuamua, $H_V=22.08$ \citep{Williams17}. For the bright population, we use the absolute magnitude of 3I/ATLAS, $H_V=17.1$ \citep{Hui2026}. We choose to neglect applying the absolute magnitude of 2I/Borisov, $H_V = 18.6$ \citep{Jewitt_2020}, as it falls between the values for 1I and 3I. It should be noted that active, cometary objects will have a range of absolute magnitudes as they get brighter closer to the star; however, we choose to use the absolute magnitude at discovery as a simplification, given the uncertainty in the size-frequency distribution of interstellar objects, and since we aim to model intrinsically bright, asteroid-like objects with this method. We account for cometary brightening laws in Section \ref{subsec: ISO_BrighteningComets_Method}.
Once we have calculated the magnitude at each position throughout each objects trajectory for both the dim and bright populations, we find the positions at which each object reaches a certain limiting magnitude, if at all, and calculate the apparent sky motion at those points. For the purposes of this calculation, we place the Earth at a true anomaly of $f_\oplus=0.0$ for both the apparent magnitude and sky motion calculations. 

\begin{table}
\begin{center}
\begin{tabular}{ccc}
\hline
\multicolumn{1}{c}{Magnitude} & \multicolumn{1}{c}{Median $d\theta/dt$ [$^\circ$/d]} & \multicolumn{1}{c}{Median $d\theta/dt$ [$^\circ$/d]}\\
\multicolumn{1}{c}{ $m_V$ } & \multicolumn{1}{c}{Dim ($H_V=22.08$)} & \multicolumn{1}{c}{Bright ($H_V=17.1$)}\\
\hline
\underline{1.2 AU}& --- & ---\\
17.0 & {1.41} & {1.57} \\
19.0 & {1.56} & {1.18} \\
21.0 & {1.65} & {0.69} \\
23.0 & {1.40} & {0.35} \\
25.0 & {0.92} & {0.17} \\
27.0 & {0.49} & {0.08} \\
\hline
\underline{3.0 AU}& --- & ---\\
17.0 & {1.34} & {1.31} \\
19.0 & {1.45} & {0.92} \\
21.0 & {1.46} & {0.57} \\
23.0 & {1.12} & {0.33} \\
25.0 & {0.73} & {0.17} \\
27.0 & {0.44} & {0.08} \\
\hline
\underline{5.0 AU}& --- & ---\\
17.0 & {1.32} & {1.29} \\
19.0 & {1.44} & {0.89} \\
21.0 & {1.44} & {0.55} \\
23.0 & {1.09} & {0.32} \\
25.0 & {0.71} & {0.18} \\
27.0 & {0.42} & {0.09} \\
\hline
\end{tabular}
\caption{Median sky motions of asteroid-like interstellar objects in 1.2 AU, 3.0 AU, and 5.0 AU spheres at limiting apparent v-band magnitudes from the distributions in Figure \ref{fig:ALL_limit_motions}.}
\label{table:all_median_motions} 
\end{center}
\end{table}

\begin{figure}
    \centering
    \includegraphics[width=\linewidth]{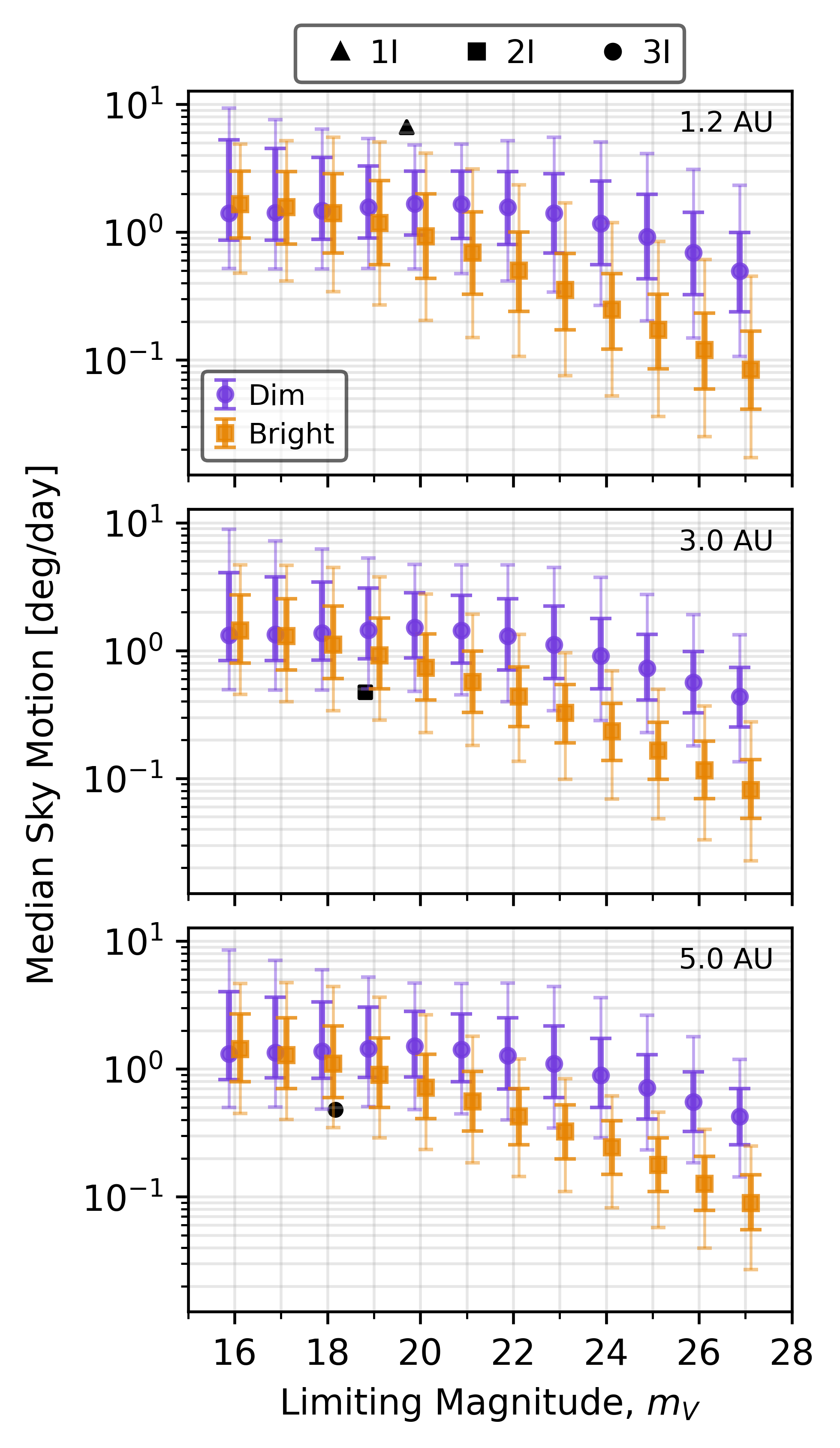}
    \caption{Summary statistics showing the median, 1-sigma, and 2-sigma quantiles for distributions of sky motions for the asteroid-like populations shown in Figure \ref{fig:ALL_limit_motions}. Points represent the median sky motion for a given distribution; error bars represent 1-sigma and 2-sigma quantiles, respectively. A sample of specific median values are provided in Table \ref{table:all_median_motions}. Reference points for the discovery conditions of the three known interstellar objects are plotted using the values in Table \ref{table: known_ISO_values_discovery_max}.}
    \label{fig:dimbright_stats}
\end{figure}

We plot the distributions of sky motions at certain magnitudes of our populations of interstellar objects within a 1.2, 3.0, and 5.0 AU sphere in Figure \ref{fig:ALL_limit_motions}. From these distributions, we extract the median sky motion values for both the bright and dim populations at each limiting magnitude, and tabulate them in Table \ref{table:all_median_motions}. It is important to note that we performed a resolution study and verified that the structure in the resulting sky motion distributions was independent of the true anomaly of the Earth (when assuming the Earth had zero eccentricity), confirming that the simulations were resolved.

\subsection{{Interstellar Comet Populations}} \label{subsec: ISO_BrighteningComets_Method}

\begin{table}
\begin{center}
\begin{tabular}{ccc}
\hline
\multicolumn{1}{c}{{Parameter}} & \multicolumn{1}{c}{{Value}} & \multicolumn{1}{c}{{Source}} \\
\hline
$\beta$ & {0.15} & \citep{Cook2016}\\
$b_1$ & {-0.13} & \citep{2011Sosa}\\
$b_2$  & {1.20} & \citep{2011Sosa}\\
$n_{pre}$  & {5.0} & \citep{2011Sosa}\\
$n_{post}$  & {5.0} & \citep{2011Sosa}\\
\hline
\end{tabular}
\caption{Parameters used to model cometary brightening laws.}
\label{table: brightening_params} 
\end{center}
\end{table}

\begin{figure*}
    \centering
    \includegraphics[width=\linewidth]{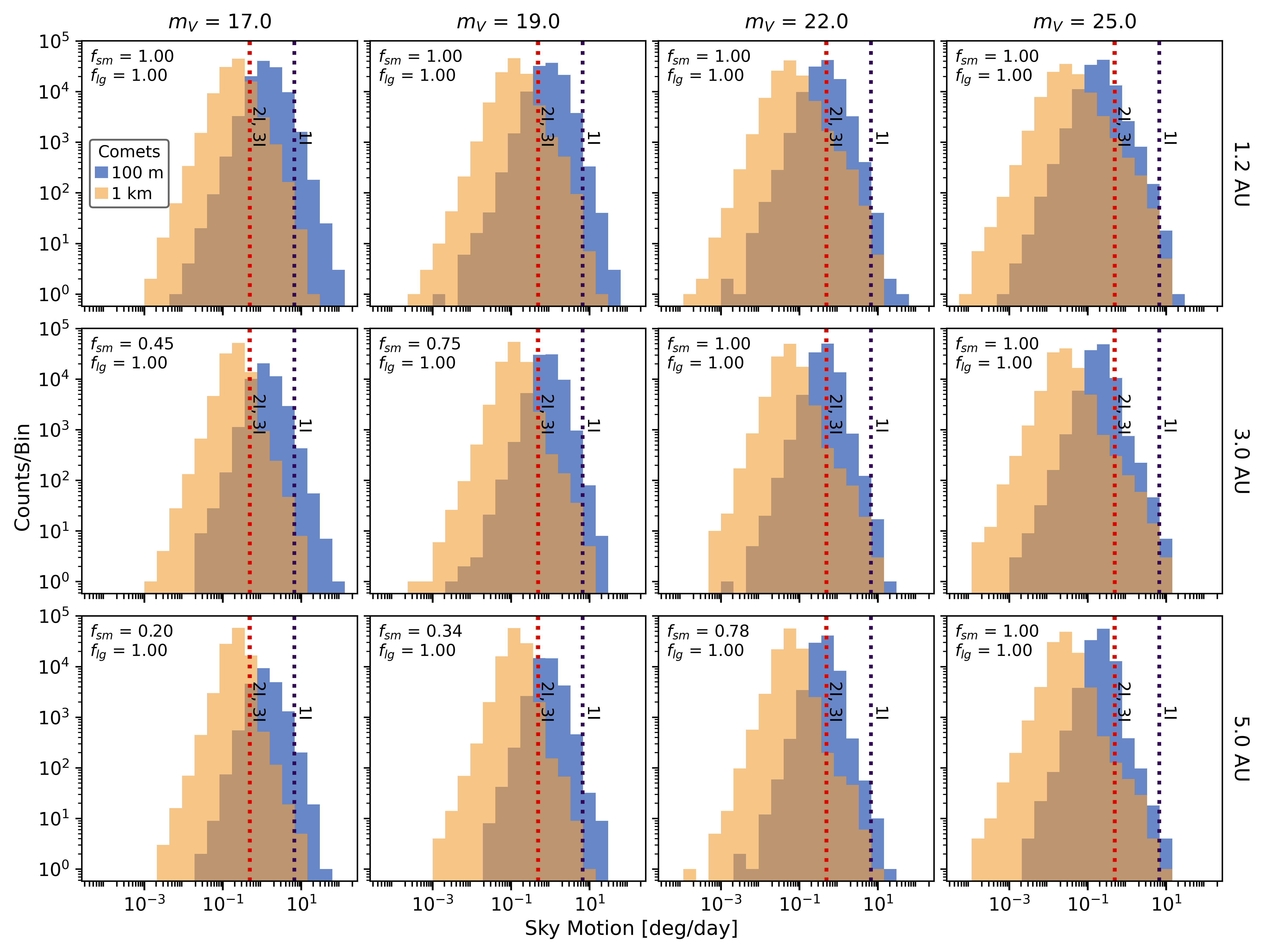}
    \caption{Similar to Figure \ref{fig:ALL_limit_motions}, but for comet-like objects, wherein the absolute and apparent magnitudes for both sets of populations are calculated using cometary brightening laws outlined in \citet{Cook2016}.
    The blue/yellow distributions represent the population of `small'/`large' objects whose absolute magnitudes were calculated using object radii of 100 m/1 km. The fraction of these 100 m/1 km objects that reach each given magnitude is denoted on the panels by $f_{sm}$/$f_{lg}$.}
    \label{fig: brightening_limit_motions}
\end{figure*}

Modeling the interstellar objects using two constant absolute magnitude values, without making an assumption about the size-frequency distributions of the population, inherently introduces a degeneracy between object size and albedo. A small, highly reflective object may have about the same absolute magnitude as a much larger, low albedo object. Additionally, these assumptions rely on the modeled interstellar object to be inactive, or lacking a visible cometary tail or coma. The presence of such activity would cause an interstellar object to increase in intrinsic brightness as it gets closer to the Sun. Two out of the three currently discovered interstellar objects have been discovered displaying cometary activity. Therefore, it is important to also account for this effect. We model this effect using the methods outlined in \citet{Cook2016}. For these two sets of populations, rather than assuming an absolute magnitude value, we instead assume a cometary nuclear radius, and calculate the cometary absolute magnitude using:
\begin{equation}
    H_c = \frac{\log_{10}(2R_c)-b_2}{b_1} \, .
    \label{eqtn: brightening_absmag}
\end{equation}
{In Equation \ref{eqtn: brightening_absmag}, $b_1$ and $b_2$ are size-brightness parameters, and $R_c$ is the radius of the comet nucleus. We set these parameters to be $b_1=-0.13$ and $b_2=1.20$ \citep{2011Sosa}, and choose two values of comet nucleus size to see the effect for small $R_c = 100$ m objects and large $R_c=1$ km objects.
Following \citet{Cook2016}, we calculate the apparent magnitude of the synthetic interstellar objects throughout their trajectory using:}
\begin{equation}
    m = H_c +2.5\bigg(\frac{n}{2}\log_{10}(d_{OS}^2)+\log_{10}(d_{OE}^2)\bigg) - 2.5\log_{10}(\gamma) \, ,
    \label{eqtn: brightening_appmag}
\end{equation}
{where $\gamma$ is a standard phase function from \citet{2010Muinonen}:}
\begin{equation}
    \gamma = (1-\beta)\Phi_1(\alpha) + \beta\Phi_1(\alpha) \, .
    \label{eqtn: brightening_phasefxn_gamma}
\end{equation}
It should be noted that $\gamma$ is derived with regard to asteroids or inactive cometary nuclei; however, \citet{Cook2016} included cometary brightening laws superposed on a nucleus with a constant absolute magnitude (Equation \ref{eqtn: brightening_absmag}) in order to calculate the apparent magnitude of a comet, as in Equation \ref{eqtn: brightening_appmag}. In Equation \ref{eqtn: brightening_phasefxn_gamma}, $\Phi_1(\alpha)$ and $\Phi_2(\alpha)$ are basis functions, given in Equation (6) of \citet{2010Muinonen}, and $\alpha$ is the phase angle described in Equation \ref{eqtn: cosine_phaseangle}. The slope parameter $\beta$ is taken to be 0.15, the typical value for objects with unmeasured phase curves \citep{Cook2016}. The variable $n$ is the photometric index of the interstellar object, which varies depending on the rate of cometary activity. Values of $n=2$ correspond to no brightening, and active comets have values ranging from $n\sim$2.5 - 7.0 \citep{2011Sosa}. For the purposes of this work, we use the same value of $n$ pre- and post-perihelion, $n_{pre}=n_{post}=5.0$. All comet brightening parameters, along with their references, are consolidated in Table \ref{table: brightening_params}.

\section{Discussion \& Conclusions} \label{sec: Discussion}

{Above, we have outlined an analytic method to calculate the on-sky motion of an object, and applied it to synthetic populations of interstellar objects with a variety of characteristics. This method allows for the near-instantaneous calculation of the on-sky motion of an object, rather than setting-up and performing a numerical integration code. Not only is the sheer calculation wall-clock time faster, this method can also save the overhead of setting up and running the numerical integration. Although JPL Horizons software can calculate the on-sky motion for a known solar system object, this method can generically be applied to any hypothetical object. Moreover, this method is generic to any observational location and any moving object orbit, although we have applied it specifically to the case of observing interstellar objects from Earth.  In this section, we discuss the resulting distributions of sky motions for these populations and how these distributions might inform future detection efforts.}

\subsection{Intrinsically Dim Population} \label{subsec: Discuss_Dim}
For our populations of near-Earth objects in a 1.2 AU sphere (top row of Figure \ref{fig:ALL_limit_motions}), the distribution of apparent sky motions for the intrinsically dim population ($H_V=22.08$) has a median value of 1.56 deg/day at an apparent magnitude of 19.0. This is $\sim$4 times slower than 1I at discovery (6.64 deg/day), and 1I was discovered at a magnitude of 19.705 \citep{Williams17}. Interestingly, this places 1I above the 2-sigma limits of our dim population distribution at $m_V=19.0$, as shown in Figure \ref{fig:dimbright_stats}. However, only a fraction (56\%) of these dim objects reach the conditions at which we serendipitously discovered 1I. Even fewer of these dim population objects, $\sim$20\% or less, reach magnitudes brighter than that, as shown by the fraction of dim objects that reach a magnitude of 17.0, labeled `$f_D$' in the upper corners of the left column of panels in Figure \ref{fig:ALL_limit_motions}. However, if smaller interstellar objects have a larger spatial number density, then the probability of detecting one of these objects would increase due to a larger population, but these objects would still be moving significantly faster, as they need to be close to Earth to be bright enough to be detected at their size. The same is true for the inner solar system objects in the 3.0 AU model sphere and the Jupiter-distance objects in the 5.0 AU model sphere. This effect could also imply that a large spatial number density of 1I-like objects is required for the detection of one such object, though we do not aim to quantify such number densities in this work.
At fainter magnitudes, as in the right columns of panels in Figure \ref{fig:ALL_limit_motions}, the objects would appear to be moving slower on the sky due to their distance from Earth being larger, but these speeds at which they would be easier to link correspond with magnitudes too faint for our all-sky surveys. For example, the NSF-DOE Vera C. Rubin Observatory, funded by the U.S. National Science Foundation and the U.S. Department of Energy's Office of Science, Legacy Survey of Space and Time (LSST) has a single-epoch r-band limiting magnitude of 24.5. This suggests that the intrinsically dim interstellar objects will be moving faster when they reach this limiting magnitude. This is because faint objects must be closer to Earth to be bright and detectable, and thus have faster median on-sky speeds (Table \ref{table:all_median_motions}). Additionally, interstellar objects are found at higher on-sky speeds for populations with steeper size-frequency distributions (see Figure 11 of \citet{Dorsey2025}).

\subsection{Intrinsically Bright Population} \label{subsec: Discuss_Bright}

For our intrinsically bright asteroid-like population, we assume an absolute magnitude value of $H_V=17.1$, the constrained absolute magnitude of 3I \citep{Hui2026}. This population reaches each apparent magnitude at farther distances than the intrinsically dim population. Thus, the distribution shows a systematic shift towards slower values at each limiting magnitude value, since sky motion scales as $d\theta/dt \sim 1/d$ (Equation \ref{eqtn: skymotion_vectoreqtn}). It is interesting to note that the bright distributions encompass the dim distributions in the $m_V=17.0$ and $m_V=19.0$ panels in Figure \ref{fig:ALL_limit_motions}. This could be partially due to the fact that a subset of the faster dim objects do not get close enough to the Earth to reach these magnitudes. However, we encourage further investigation in order to disentangle the cause for the overlapping distributions at these apparent magnitudes.

The two known interstellar comets 2I and 3I were discovered at apparent magnitudes of 18.815 and 18.170, at heliocentric distances of 3.00 and 4.51 AU, and with on-sky speeds of 0.47 and 0.48 deg/day, respectively. The median sky motion value is 1.31 deg/day for intrinsically bright asteroid-like objects in a 3.0 AU sphere when they reach a magnitude value of 17.0 (Table \ref{table:all_median_motions}), and 0.92 deg/day when they reach a magnitude of 19.0. The median values for bright asteroid populations in the 5.0 AU sphere at magnitudes 17.0 and 19.0 are similar to the 3.0 AU sphere, as shown in Table \ref{table:all_median_motions}. From this, it can be inferred that the known interstellar comets 2I and 3I were relatively slower than our modeled bright interstellar asteroids. More specifically, in the 5.0 AU model sphere, we find that 73.83\% of the bright interstellar object population, when they reach a limiting magnitude of 19.0, have speeds faster than that of 3I at discovery. Meanwhile, 51.95\% of the dim interstellar object population at magnitude 19.0 have speeds faster than 3I at discovery. 

\subsection{Comet Brightening Objects} \label{subsec: comet_brightening}

We apply cometary brightening laws to two additional sets of synthetic populations: one with assumed cometary nuclear radii of $R_c=100$ m, and the other with assumed radii of $R_c=1$ km. Given the uncertainty in the size-frequency distribution of interstellar objects, and their cometary brightening laws \citep{Cook2016}, we approximate active comets as bright point sources, and do not model their tail morphology. This is further justified as a simplification because 3I was essentially a point source at discovery \citep{Denneau2025, Seligman2025}. We then follow a similar calculation as for the asteroid-like dim and bright populations, but calculate the cometary absolute magnitude and apparent magnitude of these comet-like objects using Equations \ref{eqtn: brightening_absmag}-\ref{eqtn: brightening_phasefxn_gamma}, and the parameters listed in Table \ref{table: brightening_params}. The resulting distributions of sky motions for these populations, also generated within 1.2, 3.0, and 5.0 AU heliocentric spheres, are given in Figure \ref{fig: brightening_limit_motions}. We find that the bulk motions of these populations show similar trends, but are generally slower than those of the asteroid-like populations in Figure \ref{fig:ALL_limit_motions}. This is largely due to their rapidly increasing brightness making them detectable at larger distances, and thus appearing to move slower across the sky. For our modeled comet populations in the 5.0 AU sphere, we find that 83.94\% of our 100 m comets have motions faster than the discovery motion of 3I (0.48 deg/day) when they reach an apparent magnitude of 19.0. On the other hand, only 3.59\% of the 1 km comets in the 5.0 AU sphere have motions faster than 3I when they reach $m_V=19.0$. Observations of 3I have measured a nuclear radius of 1.3 km \citep{Hui2026}, which, compared to our 1 km distributions, imply that 3I is relatively fast for an interstellar comet of its size.

\subsection{Implications for Interstellar Object Detection} \label{subsec: ISO_detection_ability}
As of writing, only three interstellar objects have been discovered passing through our Solar System on hyperbolic trajectories, though the detection of this trio of interstellar objects implies that we should be able to find more, if we know how to look. First, we find that these interstellar objects move relatively quickly across our sky, and appear faster when they get brighter and more easily detectable, which may increase difficulty in linking them.

The intrinsic brightness of an interstellar object affects the distance at which it reaches various limiting magnitudes, is thus indirectly responsible for its on-sky speed. Comet-like objects with bright tails, such as 2I and 3I, will be brighter at further distances, and thus easier to detect when they are moving slower across the sky. Meanwhile asteroid-like objects, such as 1I, will need to get much closer to the Earth in order to be bright enough to detect, and will thus appear to be moving much faster. However, only a fraction of these dim objects get close enough to the Earth to reach an apparent magnitude of 19.0, $\sim$56\%, $\sim$19\%, and $\sim$8\% in the 1.2, 3.0, and 5.0 AU spheres, respectively.

The absolute magnitudes of interstellar objects depends significantly on the size of these objects. Small objects with a higher albedo could have a similar absolute magnitude as a large object with a low albedo. However, in this work we do not attempt to make assumptions about the size-frequency distribution of the interstellar objects in our synthetic populations. Rather, our asteroid-like populations can be thought of as a set of orbits on which we simulate an object with the intrinsic brightness properties to 1I/`Oumuamua, and the same set of orbits on which we simulate an object with the intrinsic brightness properties as 3I/ATLAS. Similarly, our cometary populations can be thought of as a set of orbits on which we simulate an object with a cometary nucleus with a radius of 100 m, and the same set of orbits on which we simulate an object with a cometary nucleus of 1 km radius.

Another factor for detectability, which is not taken into account in this work, is the position of objects relative to the Sun. We calculate the apparent rate of sky motion; however, objects at lower absolute magnitudes could be located on the opposite side of the Sun while they are both bright enough to be detected and are moving at slower on-sky speeds. In such situations, the interstellar objects would in practice be unobservable from the Earth. This effect would render some of the slower objects in our distribution undetectable. In this work, we intentionally do not include survey-specific effects. However, eliminating the interstellar objects at opposition would only increase the fraction moving at high rate of sky motions --- essentially by removing those objects at further distances preferentially. It is also possible that trailing loss impedes the discovery of these objects. Trailing loss was included in the detectability criteria of \citet{MarcetaSeligman2023}. A detailed quantification of the effect of trailing loss on the detectability of interstellar objects is outside the scope of this work, but would provide a useful contribution to the field.

It has been announced that LSST will not alert for objects moving faster than 10 deg/day\footnote{\url{https://dmtn-199.lsst.io/v/wmwv-patch-1/DMTN-199.pdf}}. However, in Table \ref{table: known_ISO_values_discovery_max}, we identify that 1I was detected moving near this limit at 6.64 deg/day on October 19, 2017, and was moving beyond this limit at 12.21 deg/day just a few days prior. This alone puts us at risk of missing out on detecting, linking, and studying these interstellar asteroids. 

Notably, the distributions of sky motions for each of our populations largely fall within the restricted velocity-search thresholds of the Near-Earth Object Surveyor mission, which ranges from 0.008-8.000 deg/day \citep{Mainzer2023}. Although we do not implement any linking algorithm analysis in this work, this lays the foundation for future analysis of the detection efficiency of interstellar objects for various all-sky surveys.

Overall, we conclude that the rapid motion of interstellar objects across our sky may pose challenges in detecting these objects, especially in the near-Earth regimes when these objects reach apparent magnitudes bright enough to detect with our surveys. However, prerecovery observations of interstellar objects 2I and 3I \citep{Ye2020,Feinstein2025,Martinez-Palomera2025,Chandler2025,Ye2025}, which had bright cometary tails, invite a promising ability to find these objects in our existing data when they are at farther distances, and moving slower across the sky. 

\section*{Acknowledgements}

CEW and DZS acknowledge funding support from JWST GO 5959, which was provided by NASA through a grant from the Space Telescope Science Institute. We thank James Wray, Tessa Frincke, Adina Feinstein, Atsuhiro Yaginuma, and Madison Brady for useful conversations. We thank the anonymous reviewer for insightful comments and constructive suggestions that strengthened the scientific content of this manuscript.

\section*{Data Availability}
 
Python scripts used for analysis and figure generation, as well as Mathematica notebooks used for equation analysis will be available once published on GitHub at this link:
\href{https://github.com/cassidywalker/ISO_SkyMotion.git}{\textcolor{black}\faGithub}

\textit{Software}: \texttt{numpy} \citep{2020Harris_NumPy}, \texttt{matplotlib} \citep{2007Hunter_Matplotlib}, \texttt{scipy} \citep{2020Virtanen_SciPy}, \texttt{pandas} \citep{2011Mckinney_pandas}.

\bibliographystyle{mnras}
\bibliography{example} 

@ARTICLE{Denneau2025,
       author = {{Denneau}, L. and {Siverd}, R. and {Tonry}, J. and {Weiland}, H. and {Erasmus}, N. and {Fitzsimmons}, A. and {Robinson}, J.},
        title = "{3I/ATLAS = C/2025 N1 (ATLAS)}",
      journal = {MPEC},
         year = 2025,
        month = jul,
       number = {2025-N12}
}

@ARTICLE{Seligman2025,
       author = {{Seligman}, Darryl Z. and {Micheli}, Marco and {Farnocchia}, Davide and {Denneau}, Larry and {Noonan}, John W. and {Hsieh}, Henry H. and {Santana-Ros}, Toni and {Tonry}, John and {Auchettl}, Katie and {Conversi}, Luca and {Devog{\`e}le}, Maxime and {Faggioli}, Laura and {Feinstein}, Adina D. and {Fenucci}, Marco and {Ferrais}, Marin and {Frincke}, Tessa and {Gillon}, Michael and {Hainaut}, Olivier R. and {Hart}, Kyle and {Hoffman}, Andrew and {Holt}, Carrie E. and {Hoogendam}, Willem B. and {Huber}, Mark E. and {Jehin}, Emmanuel and {Kareta}, Theodore and {Keane}, Jacqueline V. and {Kelley}, Michael S.~P. and {Lister}, Tim and {Mandt}, Kathleen and {Manfroid}, Jean and {Mar{\v{c}}eta}, Du{\v{s}}an and {Meech}, Karen J. and {Amine Miftah}, Mohamed and {Morgan}, Marvin and {Oca{\~n}a}, Francisco and {Pe{\~n}a-Asensio}, Eloy and {Shappee}, Benjamin J. and {Siverd}, Robert J. and {Taylor}, Aster G. and {Tucker}, Michael A. and {Wainscoat}, Richard and {Weryk}, Robert and {Wray}, James J. and {Yaginuma}, Atsuhiro and {Yang}, Bin and {Ye}, Quanzhi and {Zhang}, Qicheng},
        title = "{Discovery and Preliminary Characterization of a Third Interstellar Object: 3I/ATLAS}",
      journal = {\apjl},
         year = 2025,
        month = aug,
       volume = {989},
       number = {2},
          eid = {L36},
        pages = {L36},
          doi = {10.3847/2041-8213/adf49a},
archivePrefix = {arXiv},
       eprint = {2507.02757},
 primaryClass = {astro-ph.EP},
       adsurl = {https://ui.adsabs.harvard.edu/abs/2025ApJ...989L..36S}
}

@ARTICLE{Williams17,
       author = {{Williams}, G.~V. and {Sato}, H. and {Sarneczky}, K. and {Wainscoat}, R. and {Woodworth}, D. and {Meech}, K.},
        title = "{Minor Planets 2017 SN\_33 and 2017 U1}",
      journal = {Central Bureau Electronic Telegrams},
         year = 2017,
        month = oct,
       volume = {4450},
        pages = {1},
       adsurl = {https://ui.adsabs.harvard.edu/abs/2017CBET.4450....1W}
}

@ARTICLE{borisov_2I_cbet,
       author = {{Borisov}, G. and {Durig}, D. T. and {Sato}, H. and {Birtwhistle}, P. and {Chen}, T. and {Green}, D. W. E. and {Bacci}, P. and {Maestripieri}, M. and {Nakano}, S.},
        title = "{Comet C/2019 Q4 (Borisov)}",
      journal = {Central Bureau Electronic Telegrams},
         year = "2019",
        month = sep,
       volume = {4666},
        pages = {1}
}

@ARTICLE{Meech2017,
   author = {{Meech}, K.~J. and {Weryk}, R. and {Micheli}, M. and {Kleyna}, J.~T. and 
	{Hainaut}, O.~R. and {Jedicke}, R. and {Wainscoat}, R.~J. and 
	{Chambers}, K.~C. and {Keane}, J.~V. and {Petric}, A. and {Denneau}, L. and 
	{Magnier}, E. and {Berger}, T. and {Huber}, M.~E. and {Flewelling}, H. and 
	{Waters}, C. and {Schunova-Lilly}, E. and {Chastel}, S.},
    title = "{A brief visit from a red and extremely elongated interstellar asteroid}",
  journal = {Nature},
     year = 2017,
    month = dec,
   volume = 552,
    pages = {378-381},
      doi = {10.1038/nature25020},
   adsurl = {https://ui.adsabs.harvard.edu/abs/2017Natur.552..378M}
}

@ARTICLE{Ye2017,
   author = {{Ye}, Q.-Z. and {Zhang}, Q. and {Kelley}, M.~S.~P. and {Brown}, P.~G.
	},
    title = "{1I/2017 U1 (`Oumuamua) is Hot: Imaging, Spectroscopy, and Search of Meteor Activity}",
  journal = {\apjl},
archivePrefix = "arXiv",
   eprint = {1711.02320},
 primaryClass = "astro-ph.EP",
     year = 2017,
    month = dec,
   volume = 851,
      eid = {L5},
    pages = {L5},
      doi = {10.3847/2041-8213/aa9a34},
   adsurl = {http://adsabs.harvard.edu/abs/2017ApJ...851L...5Y}
}

@ARTICLE{Jewitt2017,
   author = {{Jewitt}, D. and {Luu}, J. and {Rajagopal}, J. and {Kotulla}, R. and 
	{Ridgway}, S. and {Liu}, W. and {Augusteijn}, T.},
    title = "{Interstellar Interloper 1I/2017 U1: Observations from the NOT and WIYN Telescopes}",
  journal = {ApJL},
archivePrefix = "arXiv",
   eprint = {1711.05687},
 primaryClass = "astro-ph.EP",
     year = 2017,
    month = dec,
   volume = 850,
      eid = {L36},
    pages = {L36},
      doi = {10.3847/2041-8213/aa9b2f},
   adsurl = {http://adsabs.harvard.edu/abs/2017ApJ...850L..36J}
}

@article{Trilling2018,
  title={Spitzer observations of interstellar object 1I/‘Oumuamua},
  author={Trilling, David E and Mommert, Michael and Hora, Joseph L and Farnocchia, Davide and Chodas, Paul and Giorgini, Jon and Smith, Howard A and Carey, Sean and Lisse, Carey M and Werner, Michael and others},
  journal={\aj},
  volume={156},
  number={6},
  pages={261},
  year={2018},
  publisher={IOP Publishing}
}

@ARTICLE{Jewitt2019b,
       author = {{Jewitt}, David and {Luu}, Jane},
        title = "{Initial Characterization of Interstellar Comet 2I/2019 Q4 (Borisov)}",
      journal = {\apjl},
         year = "2019",
        month = "Dec",
       volume = {886},
       number = {2},
          eid = {L29},
        pages = {L29},
          doi = {10.3847/2041-8213/ab530b},
archivePrefix = {arXiv},
       eprint = {1910.02547},
 primaryClass = {astro-ph.EP},
       adsurl = {https://ui.adsabs.harvard.edu/abs/2019ApJ...886L..29J}
}

@article{Jewitt_2020,
   title={Outburst and Splitting of Interstellar Comet 2I/Borisov},
   volume={896},
   ISSN={2041-8213},
   url={http://dx.doi.org/10.3847/2041-8213/ab99cb},
   DOI={10.3847/2041-8213/ab99cb},
   number={2},
   journal={The Astrophysical Journal Letters},
   publisher={American Astronomical Society},
   author={Jewitt, David and Kim, Yoonyoung and Mutchler, Max and Weaver, Harold and Agarwal, Jessica and Hui, Man-To},
   year={2020},
   month=June, pages={L39} }

@ARTICLE{2011Sosa,
       author = {{Sosa}, Andrea and {Fern{\'a}ndez}, Julio A.},
        title = "{Masses of long-period comets derived from non-gravitational effects - analysis of the computed results and the consistency and reliability of the non-gravitational parameters}",
      journal = {\mnras},
         year = 2011,
        month = sep,
       volume = {416},
       number = {1},
        pages = {767-782},
          doi = {10.1111/j.1365-2966.2011.19111.x},
       adsurl = {https://ui.adsabs.harvard.edu/abs/2011MNRAS.416..767S}
}

@ARTICLE{2010Muinonen,
       author = {{Muinonen}, Karri and {Belskaya}, Irina N. and {Cellino}, Alberto and {Delb{\`o}}, Marco and {Levasseur-Regourd}, Anny-Chantal and {Penttil{\"a}}, Antti and {Tedesco}, Edward F.},
        title = "{A three-parameter magnitude phase function for asteroids}",
      journal = {\icarus},
         year = 2010,
        month = oct,
       volume = {209},
       number = {2},
        pages = {542-555},
          doi = {10.1016/j.icarus.2010.04.003},
       adsurl = {https://ui.adsabs.harvard.edu/abs/2010Icar..209..542M}
}

@ARTICLE{Fitzsimmons2019,
       author = {{Fitzsimmons}, Alan and {Hainaut}, Olivier and {Meech}, Karen J. and {Jehin}, Emmanuel and {Moulane}, Youssef and {Opitom}, Cyrielle and {Yang}, Bin and {Keane}, Jacqueline V. and {Kleyna}, Jan T. and {Micheli}, Marco and {Snodgrass}, Colin},
        title = "{Detection of CN Gas in Interstellar Object 2I/Borisov}",
      journal = {\apjl},
         year = 2019,
        month = nov,
       volume = {885},
       number = {1},
          eid = {L9},
        pages = {L9},
          doi = {10.3847/2041-8213/ab49fc},
archivePrefix = {arXiv},
       eprint = {1909.12144},
 primaryClass = {astro-ph.EP},
       adsurl = {https://ui.adsabs.harvard.edu/abs/2019ApJ...885L...9F}
}

@ARTICLE{Ye2019,
       author = {{Ye}, Quanzhi and {Kelley}, Michael S.~P. and {Bolin}, Bryce T. and {Bodewits}, Dennis and {Farnocchia}, Davide and {Masci}, Frank J. and {Meech}, Karen J. and {Micheli}, Marco and {Weryk}, Robert and {Bellm}, Eric C. and {Christensen}, Eric and {Dekany}, Richard and {Delacroix}, Alexandre and {Graham}, Matthew J. and {Kulkarni}, Shrinivas R. and {Laher}, Russ R. and {Rusholme}, Ben and {Smith}, Roger M.},
        title = "{Pre-discovery Activity of New Interstellar Comet 2I/Borisov beyond 5 au}",
      journal = {\aj},
         year = 2020,
        month = feb,
       volume = {159},
       number = {2},
          eid = {77},
        pages = {77},
          doi = {10.3847/1538-3881/ab659b},
archivePrefix = {arXiv},
       eprint = {1911.05902},
 primaryClass = {astro-ph.EP},
       adsurl = {https://ui.adsabs.harvard.edu/abs/2020AJ....159...77Y}
}

@article{yang2021,
	Author = {Yang, Bin and Li, Aigen and Cordiner, Martin A. and Chang, Chin-Shin and Hainaut, Olivier R. and Williams, Jonathan P. and Meech, Karen J. and Keane, Jacqueline V. and Villard, Eric},
	Da = {2021/03/30},
	Doi = {10.1038/s41550-021-01336-w},
	Id = {Yang2021},
	Isbn = {2397-3366},
	Journal = {Nature Astronomy},
	Title = {Compact pebbles and the evolution of volatiles in the interstellar comet 2I/Borisov},
	Ty = {JOUR},
	Url = {https://doi.org/10.1038/s41550-021-01336-w},
	Year = {2021}}

@ARTICLE{Kim2020,
       author = {{Kim}, Yoonyoung and {Jewitt}, David and {Mutchler}, Max and {Agarwal}, Jessica and {Hui}, Man-To and {Weaver}, Harold},
        title = "{Coma Anisotropy and the Rotation Pole of Interstellar Comet 2I/Borisov}",
      journal = {\apjl},
         year = 2020,
        month = jun,
       volume = {895},
       number = {2},
          eid = {L34},
        pages = {L34},
          doi = {10.3847/2041-8213/ab9228},
archivePrefix = {arXiv},
       eprint = {2005.02468},
 primaryClass = {astro-ph.EP},
       adsurl = {https://ui.adsabs.harvard.edu/abs/2020ApJ...895L..34K}
}

@ARTICLE{McKay2020,
       author = {{McKay}, Adam J. and {Cochran}, Anita L. and {Dello Russo}, Neil and {DiSanti}, Michael A.},
        title = "{Detection of a Water Tracer in Interstellar Comet 2I/Borisov}",
      journal = {\apjl},
         year = 2020,
        month = jan,
       volume = {889},
       number = {1},
          eid = {L10},
        pages = {L10},
          doi = {10.3847/2041-8213/ab64ed},
archivePrefix = {arXiv},
       eprint = {1910.12785},
 primaryClass = {astro-ph.EP},
       adsurl = {https://ui.adsabs.harvard.edu/abs/2020ApJ...889L..10M}
}

@ARTICLE{Guzik2020,
       author = {{Guzik}, Piotr and {Drahus}, Micha{\l} and {Rusek}, Krzysztof and
         {Waniak}, Wac{\l}aw and {Cannizzaro}, Giacomo and
         {Pastor-Marazuela}, In{\'e}s},
        title = "{Initial characterization of interstellar comet 2I/Borisov}",
      journal = {Nature Astronomy},
         year = "2020",
        month = "Jan",
       volume = {4},
        pages = {53-57},
          doi = {10.1038/s41550-019-0931-8},
archivePrefix = {arXiv},
       eprint = {1909.05851},
 primaryClass = {astro-ph.EP},
       adsurl = {https://ui.adsabs.harvard.edu/abs/2020NatAs...4...53G}
}

@ARTICLE{Hui2020,
       author = {{Hui}, Man-To and {Ye}, Quan-Zhi and {F{\"o}hring}, Dora and {Hung}, Denise and {Tholen}, David J.},
        title = "{Physical Characterization of Interstellar Comet 2I/2019 Q4 (Borisov)}",
      journal = {\aj},
         year = 2020,
        month = aug,
       volume = {160},
       number = {2},
          eid = {92},
        pages = {92},
          doi = {10.3847/1538-3881/ab9df8},
       adsurl = {https://ui.adsabs.harvard.edu/abs/2020AJ....160...92H}
}

@ARTICLE{Cremonese2020,
       author = {{Cremonese}, G. and {Fulle}, M. and {Cambianica}, P. and {Munaretto}, G. and {Capria}, M.~T. and {La Forgia}, F. and {Lazzarin}, M. and {Migliorini}, A. and {Boschin}, W. and {Milani}, G. and {Aletti}, A. and {Arlic}, G. and {Bacci}, P. and {Bacci}, R. and {Bryssinck}, E. and {Carosati}, D. and {Castellano}, D. and {Buzzi}, L. and {Di Rubbo}, S. and {Facchini}, M. and {Guido}, E. and {Kugel}, F. and {Ligustri}, R. and {Maestripieri}, M. and {Mantero}, A. and {Nicolas}, J. and {Ochner}, P. and {Perrella}, C. and {Trabatti}, R. and {Valvasori}, A.},
        title = "{Dust Environment Model of the Interstellar Comet 2I/Borisov}",
      journal = {\apjl},
         year = 2020,
        month = apr,
       volume = {893},
       number = {1},
          eid = {L12},
        pages = {L12},
          doi = {10.3847/2041-8213/ab8455},
       adsurl = {https://ui.adsabs.harvard.edu/abs/2020ApJ...893L..12C}
}

@article{Marceta2023,
title = {Synthetic population of interstellar objects in the Solar System},
journal = {Astronomy and Computing},
volume = {42},
pages = {100690},
year = {2023},
issn = {2213-1337},
doi = {https://doi.org/10.1016/j.ascom.2023.100690},
url = {https://www.sciencedirect.com/science/article/pii/S2213133723000057},
author = {Dušan Marčeta}
}

@ARTICLE{Nesvorny2018,
       author = {{Nesvorn{\'y}}, David},
        title = "{Dynamical Evolution of the Early Solar System}",
      journal = {\araa},
         year = 2018,
        month = sep,
       volume = {56},
        pages = {137-174},
          doi = {10.1146/annurev-astro-081817-052028},
archivePrefix = {arXiv},
       eprint = {1807.06647},
 primaryClass = {astro-ph.EP},
       adsurl = {https://ui.adsabs.harvard.edu/abs/2018ARA&A..56..137N}
}

@ARTICLE{Tsiganis2005,
       author = {{Tsiganis}, K. and {Gomes}, R. and {Morbidelli}, A. and {Levison}, H.~F.},
        title = "{Origin of the orbital architecture of the giant planets of the Solar System}",
      journal = {\nat},
         year = "2005",
        month = "May",
       volume = {435},
       number = {7041},
        pages = {459-461},
          doi = {10.1038/nature03539},
       adsurl = {https://ui.adsabs.harvard.edu/abs/2005Natur.435..459T}
}

@ARTICLE{Hahn1999,
       author = {{Hahn}, Joseph M. and {Malhotra}, Renu},
        title = "{Orbital Evolution of Planets Embedded in a Planetesimal Disk}",
      journal = {\aj},
         year = 1999,
        month = jun,
       volume = {117},
       number = {6},
        pages = {3041-3053},
          doi = {10.1086/300891},
archivePrefix = {arXiv},
       eprint = {astro-ph/9902370},
 primaryClass = {astro-ph},
       adsurl = {https://ui.adsabs.harvard.edu/abs/1999AJ....117.3041H}
}

@ARTICLE{Gomes2004,
       author = {{Gomes}, Rodney S. and {Morbidelli}, Alessandro and {Levison}, Harold F.},
        title = "{Planetary migration in a planetesimal disk: why did Neptune stop at 30 AU?}",
      journal = {\icarus},
         year = 2004,
        month = aug,
       volume = {170},
       number = {2},
        pages = {492-507},
          doi = {10.1016/j.icarus.2004.03.011},
       adsurl = {https://ui.adsabs.harvard.edu/abs/2004Icar..170..492G}
}

@ARTICLE{Morbidelli2005,
       author = {{Morbidelli}, A. and {Levison}, H.~F. and {Tsiganis}, K. and {Gomes}, R.},
        title = "{Chaotic capture of Jupiter's Trojan asteroids in the early Solar System}",
      journal = {\nat},
         year = 2005,
        month = may,
       volume = {435},
       number = {7041},
        pages = {462-465},
          doi = {10.1038/nature03540},
       adsurl = {https://ui.adsabs.harvard.edu/abs/2005Natur.435..462M}
}

@ARTICLE{Moro2009,
   author = {{Moro-Mart{\'{\i}}n}, A. and {Turner}, E.~L. and {Loeb}, A.},
    title = "{Will the Large Synoptic Survey Telescope Detect Extra-Solar Planetesimals Entering the Solar System?}",
  journal = {\apj},
archivePrefix = "arXiv",
   eprint = {0908.3948},
 primaryClass = "astro-ph.EP",
     year = 2009,
    month = oct,
   volume = 704,
    pages = {733-742},
      doi = {10.1088/0004-637X/704/1/733},
   adsurl = {http://adsabs.harvard.edu/abs/2009ApJ...704..733M}
}

@ARTICLE{sekanina1976probability,
       author = {{Sekanina}, Z.},
        title = "{A Probability of Encounter with Interstellar Comets and the Likelihood of their Existence}",
      journal = {\icarus},
         year = 1976,
        month = jan,
       volume = {27},
       number = {1},
        pages = {123-133},
          doi = {10.1016/0019-1035(76)90189-5},
       adsurl = {https://ui.adsabs.harvard.edu/abs/1976Icar...27..123S}
}

@ARTICLE{mcglynn1989,
       author = {{McGlynn}, Thomas A. and {Chapman}, Robert D.},
        title = "{On the Nondetection of Extrasolar Comets}",
      journal = {\apjl},
         year = 1989,
        month = nov,
       volume = {346},
        pages = {L105},
          doi = {10.1086/185590},
       adsurl = {https://ui.adsabs.harvard.edu/abs/1989ApJ...346L.105M}
}

@ARTICLE{Sen1993,
       author = {{Sen}, A.~K. and {Rana}, N.~C.},
        title = "{On the missing interstellar comets}",
      journal = {\aap},
         year = 1993,
        month = aug,
       volume = {275},
        pages = {298},
       adsurl = {https://ui.adsabs.harvard.edu/abs/1993A&A...275..298S}
}

@ARTICLE{Jewitt2003,
       author = {{Jewitt}, David},
        title = "{Project Pan-STARRS and the Outer Solar System}",
      journal = {Earth Moon and Planets},
         year = 2003,
        month = jun,
       volume = {92},
       number = {1},
        pages = {465-476},
          doi = {10.1023/B:MOON.0000031961.88202.60},
       adsurl = {https://ui.adsabs.harvard.edu/abs/2003EM&P...92..465J}
}

@ARTICLE{Engelhardt2014,
       author = {{Engelhardt}, Toni and {Jedicke}, Robert and {Vere{\v{s}}}, Peter and
         {Fitzsimmons}, Alan and {Denneau}, Larry and {Beshore}, Ed and
         {Meinke}, Bonnie},
        title = "{An Observational Upper Limit on the Interstellar Number Density of Asteroids and Comets}",
      journal = {AJ},
         year = "2017",
        month = "Mar",
       volume = {153},
       number = {3},
          eid = {133},
        pages = {133},
          doi = {10.3847/1538-3881/aa5c8a},
archivePrefix = {arXiv},
       eprint = {1702.02237},
 primaryClass = {astro-ph.EP},
       adsurl = {https://ui.adsabs.harvard.edu/abs/2017AJ....153..133E}
}

@ARTICLE{Cook2016,
   author = {{Cook}, N.~V. and {Ragozzine}, D. and {Granvik}, M. and {Stephens}, D.~C.
	},
    title = "{Realistic Detectability of Close Interstellar Comets}",
  journal = {ApJ},
archivePrefix = "arXiv",
   eprint = {1607.08162},
 primaryClass = "astro-ph.EP",
     year = 2016,
    month = jul,
   volume = 825,
      eid = {51},
    pages = {51},
      doi = {10.3847/0004-637X/825/1/51},
   adsurl = {http://adsabs.harvard.edu/abs/2016ApJ...825...51C}
}

@ARTICLE{Chambers2016,
   author = {{Chambers}, K.~C. and {Magnier}, E.~A. and {Metcalfe}, N. and 
	{Flewelling}, H.~A. and {Huber}, M.~E. and {Waters}, C.~Z. and 
	{Denneau}, L. and {Draper}, P.~W. and {Farrow}, D. and {Finkbeiner}, D.~P. and 
	{Holmberg}, C. and {Koppenhoefer}, J. and {Price}, P.~A. and 
	{Rest}, A. and {Saglia}, R.~P. and {Schlafly}, E.~F. and {Smartt}, S.~J. and 
	{Sweeney}, W. and {Wainscoat}, R.~J. and {Burgett}, W.~S. and 
	{Chastel}, S. and {Grav}, T. and {Heasley}, J.~N. and {Hodapp}, K.~W. and 
	{Jedicke}, R. and {Kaiser}, N. and {Kudritzki}, R.-P. and {Luppino}, G.~A. and 
	{Lupton}, R.~H. and {Monet}, D.~G. and {Morgan}, J.~S. and {Onaka}, P.~M. and 
	{Shiao}, B. and {Stubbs}, C.~W. and {Tonry}, J.~L. and {White}, R. and 
	{Ba{\~n}ados}, E. and {Bell}, E.~F. and {Bender}, R. and {Bernard}, E.~J. and 
	{Boegner}, M. and {Boffi}, F. and {Botticella}, M.~T. and {Calamida}, A. and 
	{Casertano}, S. and {Chen}, W.-P. and {Chen}, X. and {Cole}, S. and 
	{Deacon}, N. and {Frenk}, C. and {Fitzsimmons}, A. and {Gezari}, S. and 
	{Gibbs}, V. and {Goessl}, C. and {Goggia}, T. and {Gourgue}, R. and 
	{Goldman}, B. and {Grant}, P. and {Grebel}, E.~K. and {Hambly}, N.~C. and 
	{Hasinger}, G. and {Heavens}, A.~F. and {Heckman}, T.~M. and 
	{Henderson}, R. and {Henning}, T. and {Holman}, M. and {Hopp}, U. and 
	{Ip}, W.-H. and {Isani}, S. and {Jackson}, M. and {Keyes}, C.~D. and 
	{Koekemoer}, A.~M. and {Kotak}, R. and {Le}, D. and {Liska}, D. and 
	{Long}, K.~S. and {Lucey}, J.~R. and {Liu}, M. and {Martin}, N.~F. and 
	{Masci}, G. and {McLean}, B. and {Mindel}, E. and {Misra}, P. and 
	{Morganson}, E. and {Murphy}, D.~N.~A. and {Obaika}, A. and 
	{Narayan}, G. and {Nieto-Santisteban}, M.~A. and {Norberg}, P. and 
	{Peacock}, J.~A. and {Pier}, E.~A. and {Postman}, M. and {Primak}, N. and 
	{Rae}, C. and {Rai}, A. and {Riess}, A. and {Riffeser}, A. and 
	{Rix}, H.~W. and {R{\"o}ser}, S. and {Russel}, R. and {Rutz}, L. and 
	{Schilbach}, E. and {Schultz}, A.~S.~B. and {Scolnic}, D. and 
	{Strolger}, L. and {Szalay}, A. and {Seitz}, S. and {Small}, E. and 
	{Smith}, K.~W. and {Soderblom}, D.~R. and {Taylor}, P. and {Thomson}, R. and 
	{Taylor}, A.~N. and {Thakar}, A.~R. and {Thiel}, J. and {Thilker}, D. and 
	{Unger}, D. and {Urata}, Y. and {Valenti}, J. and {Wagner}, J. and 
	{Walder}, T. and {Walter}, F. and {Watters}, S.~P. and {Werner}, S. and 
	{Wood-Vasey}, W.~M. and {Wyse}, R.},
    title = "{The Pan-STARRS1 Surveys}",
  journal = {arXiv e-prints},
archivePrefix = "arXiv",
   eprint = {1612.05560},
 primaryClass = "astro-ph.IM",
     year = 2016,
    month = dec,
   adsurl = {https://ui.adsabs.harvard.edu/abs/2016arXiv161205560C}
}

@ARTICLE{Trilling2017,
       author = {{Trilling}, David E. and {Robinson}, Tyler and {Roegge}, Alissa and {Chand
        ler}, Colin Orion and {Smith}, Nathan and {Loeffler}, Mark and
         {Trujillo}, Chad and {Navarro-Meza}, Samuel and {Glaspie}, Lori M.},
        title = "{Implications for Planetary System Formation from Interstellar Object 1I/2017 U1 ({\textquoteleft}Oumuamua)}",
      journal = {\apjl},
         year = "2017",
        month = "Dec",
       volume = {850},
       number = {2},
          eid = {L38},
        pages = {L38},
          doi = {10.3847/2041-8213/aa9989},
archivePrefix = {arXiv},
       eprint = {1711.01344},
 primaryClass = {astro-ph.EP},
       adsurl = {https://ui.adsabs.harvard.edu/abs/2017ApJ...850L..38T}
}

@ARTICLE{Laughlin2017,
   author = {{Laughlin}, G. and {Batygin}, K.},
    title = "{On the Consequences of the Detection of an Interstellar Asteroid}",
  journal = {Research Notes of the American Astronomical Society},
archivePrefix = "arXiv",
   eprint = {1711.02260},
 primaryClass = "astro-ph.EP",
     year = 2017,
    month = dec,
   volume = 1,
   number = 1,
      eid = {43},
    pages = {43},
      doi = {10.3847/2515-5172/aaa02b},
   adsurl = {http://adsabs.harvard.edu/abs/2017RNAAS...1a..43L}
}

@ARTICLE{Do2018,
   author = {{Do}, A. and {Tucker}, M.~A. and {Tonry}, J.},
    title = "{Interstellar Interlopers: Number Density and Origin of Oumuamua-like Objects}",
  journal = {ApJL},
archivePrefix = "arXiv",
   eprint = {1801.02821},
 primaryClass = "astro-ph.EP",
     year = 2018,
    month = mar,
   volume = 855,
      eid = {L10},
    pages = {L10},
      doi = {10.3847/2041-8213/aaae67},
   adsurl = {http://adsabs.harvard.edu/abs/2018ApJ...855L..10D}
}

@ARTICLE{MoroMartin2018i,
       author = {{Moro-Mart{\'\i}n}, Amaya},
        title = "{Origin of 1I/{\textquoteright}Oumuamua. I. An Ejected Protoplanetary Disk Object?}",
      journal = {\apj},
         year = 2018,
        month = oct,
       volume = {866},
       number = {2},
          eid = {131},
        pages = {131},
          doi = {10.3847/1538-4357/aadf34},
archivePrefix = {arXiv},
       eprint = {1810.02148},
 primaryClass = {astro-ph.EP},
       adsurl = {https://ui.adsabs.harvard.edu/abs/2018ApJ...866..131M}
}

@ARTICLE{Moro2019exOC,
       author = {{Moro-Mart{\'\i}n}, Amaya},
        title = "{Origin of 1I/{\textquoteright}Oumuamua. II. An Ejected Exo-Oort Cloud Object?}",
      journal = {\aj},
         year = 2019,
        month = feb,
       volume = {157},
       number = {2},
          eid = {86},
        pages = {86},
          doi = {10.3847/1538-3881/aafda6},
archivePrefix = {arXiv},
       eprint = {1811.00023},
 primaryClass = {astro-ph.EP},
       adsurl = {https://ui.adsabs.harvard.edu/abs/2019AJ....157...86M}
}

@ARTICLE{Hui2026,
       author = {{Hui}, Man-To and {Jewitt}, David and {Mutchler}, Max J. and {Agarwal}, Jessica and {Kim}, Yoonyoung},
        title = "{Nucleus and Postperihelion Activity of Interstellar Object 3I/ATLAS Observed by Hubble Space Telescope}",
      journal = {arXiv e-prints},
         year = 2026,
        month = jan,
          eid = {arXiv:2601.21569},
        pages = {arXiv:2601.21569},
          doi = {10.48550/arXiv.2601.21569},
archivePrefix = {arXiv},
       eprint = {2601.21569},
 primaryClass = {astro-ph.EP},
       adsurl = {https://ui.adsabs.harvard.edu/abs/2026arXiv260121569H}
}

@ARTICLE{Hoover2022,
       author = {{Hoover}, Devin J. and {Seligman}, Darryl Z. and {Payne}, Matthew J.},
        title = "{The Population of Interstellar Objects Detectable with the LSST and Accessible for In Situ Rendezvous with Various Mission Designs}",
      journal = {\psj},
         year = 2022,
        month = mar,
       volume = {3},
       number = {3},
          eid = {71},
        pages = {71},
          doi = {10.3847/PSJ/ac58fe},
archivePrefix = {arXiv},
       eprint = {2109.10406},
 primaryClass = {astro-ph.EP},
       adsurl = {https://ui.adsabs.harvard.edu/abs/2022PSJ.....3...71H}
}

@ARTICLE{Dorsey2025,
       author = {{Dorsey}, Rosemary C. and {Hopkins}, Matthew J. and {Bannister}, Michele T. and {Lawler}, Samantha M. and {Lintott}, Chris and {Parker}, Alex H. and {Forbes}, John C.},
        title = "{The Visibility of the {\={O}}tautahi─Oxford Interstellar Object Population Model in LSST}",
      journal = {\psj},
         year = 2025,
        month = sep,
       volume = {6},
       number = {9},
          eid = {214},
        pages = {214},
          doi = {10.3847/PSJ/adf8ca},
archivePrefix = {arXiv},
       eprint = {2502.16741},
 primaryClass = {astro-ph.EP},
       adsurl = {https://ui.adsabs.harvard.edu/abs/2025PSJ.....6..214D}
}

@ARTICLE{Micheli2018,
       author = {{Micheli}, Marco and {Farnocchia}, Davide and {Meech}, Karen J. and {Buie}, Marc W. and {Hainaut}, Olivier R. and {Prialnik}, Dina and {Sch{\"o}rghofer}, Norbert and {Weaver}, Harold A. and {Chodas}, Paul W. and {Kleyna}, Jan T. and {Weryk}, Robert and {Wainscoat}, Richard J. and {Ebeling}, Harald and {Keane}, Jacqueline V. and {Chambers}, Kenneth C. and {Koschny}, Detlef and {Petropoulos}, Anastassios E.},
        title = "{Non-gravitational acceleration in the trajectory of 1I/2017 U1 ('Oumuamua)}",
      journal = {\nat},
         year = 2018,
        month = jun,
       volume = {559},
        pages = {223-226},
          doi = {10.1038/s41586-018-0254-4},
       adsurl = {https://ui.adsabs.harvard.edu/abs/2018Natur.559..223M}
}

@ARTICLE{Tonry2018a,
       author = {{Tonry}, J.~L. and {Denneau}, L. and {Heinze}, A.~N. and {Stalder}, B. and {Smith}, K.~W. and {Smartt}, S.~J. and {Stubbs}, C.~W. and {Weiland}, H.~J. and {Rest}, A.},
        title = "{ATLAS: A High-cadence All-sky Survey System}",
      journal = {\pasp},
         year = 2018,
        month = jun,
       volume = {130},
       number = {988},
        pages = {064505},
          doi = {10.1088/1538-3873/aabadf},
archivePrefix = {arXiv},
       eprint = {1802.00879},
 primaryClass = {astro-ph.IM},
       adsurl = {https://ui.adsabs.harvard.edu/abs/2018PASP..130f4505T}
}

@ARTICLE{Tonry2018b,
       author = {{Tonry}, J.~L. and {Denneau}, L. and {Flewelling}, H. and {Heinze}, A.~N. and {Onken}, C.~A. and {Smartt}, S.~J. and {Stalder}, B. and {Weiland}, H.~J. and {Wolf}, C.},
        title = "{The ATLAS All-Sky Stellar Reference Catalog}",
      journal = {\apj},
         year = 2018,
        month = nov,
       volume = {867},
       number = {2},
          eid = {105},
        pages = {105},
          doi = {10.3847/1538-4357/aae386},
archivePrefix = {arXiv},
       eprint = {1809.09157},
 primaryClass = {astro-ph.IM},
       adsurl = {https://ui.adsabs.harvard.edu/abs/2018ApJ...867..105T}
}

@article{MoroMartin2019,
  title={Could 1I/’Oumuamua be an Icy Fractal Aggregate?},
  author={Moro-Mart{\'\i}n, Amaya},
  journal={\apjl},
  volume={872},
  number={2},
  pages={L32},
  year={2019},
  publisher={IOP Publishing}
}

@ARTICLE{Flekkoy19,
       author = {{Flekk{\o}y}, Eirik G. and {Luu}, Jane and {Toussaint}, Renaud},
        title = "{The Interstellar Object {\textquoteright}Oumuamua as a Fractal Dust Aggregate}",
      journal = {\apjl},
         year = 2019,
        month = nov,
       volume = {885},
       number = {2},
          eid = {L41},
        pages = {L41},
          doi = {10.3847/2041-8213/ab4f78},
archivePrefix = {arXiv},
       eprint = {1910.07135},
 primaryClass = {astro-ph.EP},
       adsurl = {https://ui.adsabs.harvard.edu/abs/2019ApJ...885L..41F}
}

@ARTICLE{Luu20,
       author = {{Luu}, Jane X. and {Flekk{\o}y}, Eirik G. and {Toussaint}, Renaud},
        title = "{'Oumuamua as a Cometary Fractal Aggregate: The ``Dust Bunny'' Model}",
      journal = {\apjl},
         year = 2020,
        month = sep,
       volume = {900},
       number = {2},
          eid = {L22},
        pages = {L22},
          doi = {10.3847/2041-8213/abafa7},
archivePrefix = {arXiv},
       eprint = {2008.10083},
 primaryClass = {astro-ph.EP},
       adsurl = {https://ui.adsabs.harvard.edu/abs/2020ApJ...900L..22L}
}

@ARTICLE{Bialy2018,
       author = {{Bialy}, Shmuel and {Loeb}, Abraham},
        title = "{Could Solar Radiation Pressure Explain {\textquoteleft}Oumuamua{\textquoteright}s Peculiar Acceleration?}",
      journal = {\apjl},
         year = 2018,
        month = nov,
       volume = {868},
       number = {1},
          eid = {L1},
        pages = {L1},
          doi = {10.3847/2041-8213/aaeda8},
archivePrefix = {arXiv},
       eprint = {1810.11490},
 primaryClass = {astro-ph.EP},
       adsurl = {https://ui.adsabs.harvard.edu/abs/2018ApJ...868L...1B}
}

@article{jackson20211i,
  title={1I/‘Oumuamua as an N2 ice fragment of an exo-Pluto surface: I. Size and Compositional Constraints},
  author={Jackson, Alan P and Desch, Steven J},
  journal={Journal of Geophysical Research: Planets},
  pages={e2020JE006706},
  year={2021},
  publisher={Wiley Online Library}
}

@article{desch20211i,
  title={1I/‘Oumuamua as an N2 ice fragment of an exo-pluto surface II: Generation of N2 ice fragments and the origin of ‘Oumuamua},
  author={Desch, Steven J and Jackson, Alan P},
  journal={Journal of Geophysical Research: Planets},
  pages={e2020JE006807},
  year={2021},
  publisher={Wiley Online Library}
}

@ARTICLE{Desch2022,
       author = {{Desch}, Steven J. and {Jackson}, Alan P.},
        title = "{Some Pertinent Issues for Interstellar Panspermia Raised after the Discovery of 1I/`Oumuamua}",
      journal = {Astrobiology},
         year = 2022,
        month = dec,
       volume = {22},
       number = {12},
        pages = {1400-1413},
          doi = {10.1089/ast.2021.0199},
       adsurl = {https://ui.adsabs.harvard.edu/abs/2022AsBio..22.1400D}
}

@ARTICLE{Sekanina2019,
       author = {{Sekanina}, Zdenek},
        title = "{Outgassing As Trigger of 1I/`Oumuamua's Nongravitational Acceleration: Could This Hypothesis Work at All?}",
      journal = {arXiv e-prints},
         year = 2019,
        month = may,
          eid = {arXiv:1905.00935},
        pages = {arXiv:1905.00935},
archivePrefix = {arXiv},
       eprint = {1905.00935},
 primaryClass = {astro-ph.EP},
       adsurl = {https://ui.adsabs.harvard.edu/abs/2019arXiv190500935S}
}

@ARTICLE{Seligman2020,
       author = {{Seligman}, Darryl and {Laughlin}, Gregory},
        title = "{Evidence that 1I/2017 U1 ('Oumuamua) was Composed of Molecular Hydrogen Ice}",
      journal = {\apjl},
         year = 2020,
        month = jun,
       volume = {896},
       number = {1},
          eid = {L8},
        pages = {L8},
          doi = {10.3847/2041-8213/ab963f},
archivePrefix = {arXiv},
       eprint = {2005.12932},
 primaryClass = {astro-ph.EP},
       adsurl = {https://ui.adsabs.harvard.edu/abs/2020ApJ...896L...8S}
}

@ARTICLE{Levine2021_h2,
       author = {{Levine}, W. Garrett and {Laughlin}, Gregory},
        title = "{Assessing the Formation of Solid Hydrogen Objects in Starless Molecular Cloud Cores}",
      journal = {\apj},
         year = 2021,
        month = may,
       volume = {912},
       number = {1},
          eid = {3},
        pages = {3},
          doi = {10.3847/1538-4357/abec85},
archivePrefix = {arXiv},
       eprint = {2103.05449},
 primaryClass = {astro-ph.EP},
       adsurl = {https://ui.adsabs.harvard.edu/abs/2021ApJ...912....3L}
}

@ARTICLE{Bergner2023,
       author = {{Bergner}, Jennifer B. and {Seligman}, Darryl Z.},
        title = "{Acceleration of 1I/`Oumuamua from radiolytically produced H$_{2}$ in H$_{2}$O ice}",
      journal = {\nat},
         year = 2023,
        month = mar,
       volume = {615},
       number = {7953},
        pages = {610-613},
          doi = {10.1038/s41586-022-05687-w},
archivePrefix = {arXiv},
       eprint = {2303.13698},
 primaryClass = {astro-ph.EP},
       adsurl = {https://ui.adsabs.harvard.edu/abs/2023Natur.615..610B}
}

@ARTICLE{Levine2021,
       author = {{Levine}, W. Garrett and {Cabot}, Samuel H.~C. and {Seligman}, Darryl and {Laughlin}, Gregory},
        title = "{Constraints on the Occurrence of 'Oumuamua-Like Objects}",
      journal = {\apj},
         year = 2021,
        month = nov,
       volume = {922},
       number = {1},
          eid = {39},
        pages = {39},
          doi = {10.3847/1538-4357/ac1fe6},
archivePrefix = {arXiv},
       eprint = {2108.11194},
 primaryClass = {astro-ph.EP},
       adsurl = {https://ui.adsabs.harvard.edu/abs/2021ApJ...922...39L}
}

@ARTICLE{Seligman2024PNAS,
       author = {{Seligman}, Darryl Z. and {Farnocchia}, Davide and {Micheli}, Marco and {Hainaut}, Olivier R. and {Hsieh}, Henry H. and {Feinstein}, Adina D. and {Chesley}, Steven R. and {Taylor}, Aster G. and {Masiero}, Joseph and {Meech}, Karen J.},
        title = "{Two distinct populations of dark comets delineated by orbits and sizes}",
      journal = {Proceedings of the National Academy of Science},
         year = 2024,
        month = dec,
       volume = {121},
       number = {51},
          eid = {e2406424121},
        pages = {e2406424121},
          doi = {10.1073/pnas.2406424121},
archivePrefix = {arXiv},
       eprint = {2412.07603},
 primaryClass = {astro-ph.EP},
       adsurl = {https://ui.adsabs.harvard.edu/abs/2024PNAS..12106424S}
}

@ARTICLE{Seligman2023b,
       author = {{Seligman}, Darryl Z. and {Farnocchia}, Davide and {Micheli}, Marco and {Vokrouhlick{\'y}}, David and {Taylor}, Aster G. and {Chesley}, Steven R. and {Bergner}, Jennifer B. and {Vere{\v{s}}}, Peter and {Hainaut}, Olivier R. and {Meech}, Karen J. and {Devogele}, Maxime and {Pravec}, Petr and {Matson}, Rob and {Deen}, Sam and {Tholen}, David J. and {Weryk}, Robert and {Rivera-Valent{\'\i}n}, Edgard G. and {Sharkey}, Benjamin N.~L.},
        title = "{Dark Comets? Unexpectedly Large Nongravitational Accelerations on a Sample of Small Asteroids}",
      journal = {\psj},
         year = 2023,
        month = feb,
       volume = {4},
       number = {2},
          eid = {35},
        pages = {35},
          doi = {10.3847/PSJ/acb697},
archivePrefix = {arXiv},
       eprint = {2212.08115},
 primaryClass = {astro-ph.EP},
       adsurl = {https://ui.adsabs.harvard.edu/abs/2023PSJ.....4...35S}
}

@ARTICLE{Farnocchia2023,
       author = {{Farnocchia}, Davide and {Seligman}, Darryl Z. and {Granvik}, Mikael and {Hainaut}, Olivier and {Meech}, Karen J. and {Micheli}, Marco and {Weryk}, Robert and {Chesley}, Steven R. and {Christensen}, Eric J. and {Koschny}, Detlef and {Kleyna}, Jan T. and {Lazzaro}, Daniela and {Mommert}, Michael and {Wainscoat}, Richard J.},
        title = "{(523599) 2003 RM: The Asteroid that Wanted to be a Comet}",
      journal = {\psj},
         year = 2023,
        month = feb,
       volume = {4},
       number = {2},
          eid = {29},
        pages = {29},
          doi = {10.3847/PSJ/acb25b},
archivePrefix = {arXiv},
       eprint = {2212.08135},
 primaryClass = {astro-ph.EP},
       adsurl = {https://ui.adsabs.harvard.edu/abs/2023PSJ.....4...29F}
}

@ARTICLE{Opitom2025,
       author = {{Opitom}, Cyrielle and {Snodgrass}, Colin and {Jehin}, Emmanuel and {Bannister}, Michele T. and {Bufanda}, Erica and {Deam}, Sophie E. and {Dorsey}, Rosemary C. and {Ferrais}, Marin and {Hmiddouch}, Said and {Knight}, Matthew M. and {Kokotanekova}, Rosita and {Leicester}, Brayden and {Marsset}, Micha{\"e}l and {Murphy}, Brian and {Okoth}, Vincent and {Ridden-Harper}, Ryan and {Vander Donckt}, Mathieu and {Ferellec}, L{\'e}a and {Hutsem{\'e}kers}, Damien and {Lippi}, Manuela and {Manfroid}, Jean and {Benkhaldoun}, Zouhair},
        title = "{Snapshot of a new interstellar comet: 3I/ATLAS has a red and featureless spectrum}",
      journal = {\mnras},
         year = 2025,
        month = nov,
       volume = {544},
       number = {1},
        pages = {L31-L36},
          doi = {10.1093/mnrasl/slaf095},
archivePrefix = {arXiv},
       eprint = {2507.05226},
 primaryClass = {astro-ph.EP},
       adsurl = {https://ui.adsabs.harvard.edu/abs/2025MNRAS.544L..31O}
}

@ARTICLE{Kareta2025,
       author = {{Kareta}, Theodore and {Champagne}, Chansey and {McClure}, Lucas and {Emery}, Joshua and {Sharkey}, Benjamin N.~L. and {Bauer}, James and {Connelley}, Michael S. and {Rayner}, John and {Thomas}, Cristina A. and {Reddy}, Vishnu and {Firgard}, Megan},
        title = "{Near-discovery Observations of Interstellar Comet 3I/ATLAS with the NASA Infrared Telescope Facility}",
      journal = {\apjl},
         year = 2025,
        month = sep,
       volume = {990},
       number = {2},
          eid = {L65},
        pages = {L65},
          doi = {10.3847/2041-8213/adfbdf},
archivePrefix = {arXiv},
       eprint = {2507.12234},
 primaryClass = {astro-ph.EP},
       adsurl = {https://ui.adsabs.harvard.edu/abs/2025ApJ...990L..65K}
}

@ARTICLE{Jewitt2025,
       author = {{Jewitt}, David and {Luu}, Jane},
        title = "{Interstellar Interloper C/2025 N1 is Active}",
      journal = {The Astronomer's Telegram},
         year = 2025,
        month = jul,
       volume = {17263},
        pages = {1},
       adsurl = {https://ui.adsabs.harvard.edu/abs/2025ATel17263....1J}
}

@ARTICLE{Yang2025,
       author = {{Yang}, Bin and {Meech}, Karen J. and {Connelley}, Michael and {Zhao}, Ruining and {Keane}, Jacqueline V.},
        title = "{Spectroscopic Characterization of Interstellar Object 3I/ATLAS: Water Ice in the Coma}",
      journal = {\apjl},
         year = 2025,
        month = oct,
       volume = {992},
       number = {1},
          eid = {L9},
        pages = {L9},
          doi = {10.3847/2041-8213/ae08a7},
archivePrefix = {arXiv},
       eprint = {2507.14916},
 primaryClass = {astro-ph.EP},
       adsurl = {https://ui.adsabs.harvard.edu/abs/2025ApJ...992L...9Y}
}

@ARTICLE{Chandler2025,
       author = {{Chandler}, Colin Orion and {Bernardinelli}, Pedro H. and {Juri{\'c}}, Mario and {Singh}, Devanshi and {Hsieh}, Henry H. and {Sullivan}, Ian and {Jones}, R. Lynne and {Kurlander}, Jacob A. and {Vavilov}, Dmitrii and {Eggl}, Siegfried and {Holman}, Matthew and {Spoto}, Federica and {Schwamb}, Megan E. and {Christensen}, Eric J. and {Beebe}, Wilson and {Roodman}, Aaron and {Lim}, Kian-Tat and {Jenness}, Tim and {Bosch}, James and {Smart}, Brianna and {Bellm}, Eric and {MacBride}, Sean and {Rawls}, Meredith L. and {Greenstreet}, Sarah and {Slater}, Colin and {Heinze}, Aren and {Ivezi{\'c}}, {\v{Z}}eljko and {Blum}, Bob and {Connolly}, Andrew and {Daues}, Gregory and {Makadia}, Rahil and {Gower}, Michelle and {Bryce Kalmbach}, J. and {Monet}, David and {Bannister}, Michele T. and {Dones}, Luke and {Dorsey}, Rosemary C. and {Fraser}, Wesley C. and {Forbes}, John C. and {Fuentes}, Cesar and {Holt}, Carrie E. and {Inno}, Laura and {Jones}, Geraint H. and {Knight}, Matthew M. and {Lintott}, Chris J. and {Lister}, Tim and {Lupton}, Robert and {Mendoza Magbanua}, Mark Jesus and {Malhotra}, Renu and {Mueller}, Beatrice E.~A. and {Murtagh}, Joseph and {Pandey}, Nitya and {Reach}, William T. and {Samarasinha}, Nalin H. and {Seligman}, Darryl Z. and {Snodgrass}, Colin and {Solontoi}, Michael and {Szab{\'o}}, Gyula M. and {White}, Ellie and {Womack}, Maria and {Young}, Leslie A. and {Allbery}, Russ and {Armellin}, Roberto and {Aubourg}, {\'E}ric and {Avdellidou}, Chrysa and {Azfar}, Farrukh and {Bauer}, James and {Bechtol}, Keith and {Belyakov}, Matthew and {Benecchi}, Susan D. and {Bertini}, Ivano and {Bolin}, Bryce T. and {Bose}, vMaitrayee and {Buchanan}, Laura E. and {Boucaud}, Alexandre and {Boufleur}, Rodrigo C. and {Boutigny}, Dominique and {Braga-Ribas}, Felipe and {Calabrese}, Daniel and {Camargo}, J.~I.~B. and {Caplar}, Neven and {Carry}, Benoit and {Carvajal}, Juan Pablo and {Choi}, Yumi and {Cowan}, Preeti and {Croft}, Steve and {{\'C}uk}, Matija and {Daruich}, Felipe and {Daubard}, Guillaume and {Davenport}, James R.~A. and {Daylan}, Tansu and {Delgado}, Jennifer and {Devillepoix}, Hadrien A.~R. and {Doherty}, Peter E. and {Donaldson}, Abbie and {Drass}, Holger and {Deppe}, Stephanie JH and {Dubois-Felsmann}, Gregory P. and {Economou}, Frossie and {Eduardo}, Marielle R. and {Farnocchia}, Davide and {Frissell}, Maxwell K. and {Fedorets}, Grigori and {Fernandes}, Maryann Benny and {Fulle}, Marco and {Gerdes}, David W. and {Gibbs}, Alex R. and {Gillan}, A. Fraser and {Guy}, Leanne P. and {Hammergren}, Mark and {Hanushevsky}, Andrew and {Hernandez}, Fabio and {Hestroffer}, Daniel and {Hopkins}, Matthew J. and {Granvik}, Mikael and {Ieva}, Simone and {Irving}, David H. and {Jannuzi}, Buell T. and {Jimenez}, David and {Ramos Gomes-J{\'u}nior}, Altair and {Juramy}, Claire and {Kahn}, Steven M. and {Kannawadi}, Arun and {Kang}, Yijung and {Kryszczy{\'n}ska}, Agnieszka and {Kotov}, Ivan and {Koumjian}, Alec and {Krughoff}, K. Simon and {Lage}, Craig and {Lange}, Travis J. and {Levine}, W. Garrett and {Li}, Zhuofu and {Licandro}, Javier and {Lin}, Hsing Wen and {Lust}, Nate B. and {Lyttle}, Ryan R. and {Mahabal}, Ashish A. and {Mahlke}, Max and {Plazas Malag{\'o}n}, Andr{\'e}s A. and {Salazar Manzano}, Luis E. and {Marc}, Moniez and {Margoti}, Giuliano and {Mar{\v{c}}eta}, Du{\v{s}}an and {Menanteau}, Felipe and {Meyers}, Joshua and {Mills}, Dave and {Morato}, Naomi and {More}, Surhud and {Morrison}, Christopher B. and {Moulane}, Youssef and {Mu{\~n}oz-Guti{\'e}rrez}, Marco A. and {M.}, Newcomer F. and {O'Connor}, Paul and {Oldag}, Drew and {Oldroyd}, William J and {O'Mullane}, William and {Opitom}, Cyrielle and {Oszkiewicz}, Dagmara and {Page}, Gary L. and {Patterson}, Jack and {Payne}, Matthew J. and {Peloton}, Julien and {Pereira}, Chrystian Luciano and {Peterson}, John R. and {Polin}, Daniel and {Pollek}, Hannah Mary Margaret and {Polen}, Rebekah and {Qiu}, Yongqiang and {Ragozzine}, Darin and {Rajagopal}, Jayadev and {van Reeven}, vWouter and {Rice}, Malena and {Ridgway}, Stephen T. and {Rivkin}, Andrew S. and {Robinson}, James E. and {Ro{\.z}ek}, Agata and {Salnikov}, Andrei and {S{\'a}nchez}, Bruno O. and {Sarid}, Gal and {Schambeau}, Charles A. and {Scolnic}, Daniel and {Schindler}, Rafe H. and {Seaman}, Robert and {Jacques}, {\v{S}}ebag and {Shaw}, Richard A. and {Shugart}, Alysha and {Sick}, Jonathan and {Siraj}, Amir and {Sitarz}, Michael C. and {Sobhani}, Shahram and {Soldahl}, Christine and {Stalder}, Brian and {Stetzler}, Steven and {Swinbank}, John D. and {Szigeti}, L{\'a}szl{\'o} and {Tauraso}, Michael and {Thornton}, Adam and {Tonietti}, Luca and {Trilling}, David E. and {Trujillo}, Chadwick A.},
        title = "{NSF-DOE Vera C. Rubin Observatory Observations of Interstellar Comet 3I/ATLAS (C/2025 N1)}",
      journal = {arXiv e-prints},
         year = 2025,
        month = jul,
          eid = {arXiv:2507.13409},
        pages = {arXiv:2507.13409},
archivePrefix = {arXiv},
       eprint = {2507.13409},
 primaryClass = {astro-ph.EP},
       adsurl = {https://ui.adsabs.harvard.edu/abs/2025arXiv250713409C}
}

@ARTICLE{Belyakov2025,
       author = {{Belyakov}, Matthew and {Fremling}, Christoffer and {Graham}, Matthew J. and {Bolin}, Bryce T. and {Kilic}, Mukremin and {Jewett}, Gracyn and {Lisse}, Carey M. and {Ingebretsen}, Carl and {Davis}, M. Ryleigh and {Wong}, Ian},
        title = "{Palomar and Apache Point Spectrophotometry of Interstellar Comet 3I/ATLAS}",
      journal = {Research Notes of the American Astronomical Society},
         year = 2025,
        month = jul,
       volume = {9},
       number = {7},
          eid = {194},
        pages = {194},
          doi = {10.3847/2515-5172/adf059},
archivePrefix = {arXiv},
       eprint = {2507.11720},
 primaryClass = {astro-ph.EP},
       adsurl = {https://ui.adsabs.harvard.edu/abs/2025RNAAS...9..194B}
}

@ARTICLE{Marcos2025,
       author = {{de la Fuente Marcos}, R. and {Alarcon}, M.~R. and {Licandro}, J. and {Serra-Ricart}, M. and {de Le{\'o}n}, J. and {de la Fuente Marcos}, C. and {Lombardi}, G. and {Tejero}, A. and {Cabrera-Lavers}, A. and {Guerra Arencibia}, S. and {Ruiz Cejudo}, I.},
        title = "{Assessing interstellar comet 3I/ATLAS with the 10.4 m Gran Telescopio Canarias and the Two-meter Twin Telescope}",
      journal = {\aap},
         year = 2025,
        month = aug,
       volume = {700},
          eid = {L9},
        pages = {L9},
          doi = {10.1051/0004-6361/202556439},
archivePrefix = {arXiv},
       eprint = {2507.12922},
 primaryClass = {astro-ph.EP},
       adsurl = {https://ui.adsabs.harvard.edu/abs/2025A&A...700L...9D}
}

@ARTICLE{Alarcon2025,
       author = {{Alarcon}, Miguel R. and {Serra-Ricart}, Miquel and {Licandro}, Javier and {Arencibia}, Sergio Guerra and {Ruiz Cejudo}, Ignacio and {Trujillo}, Ignacio},
        title = "{Deep g'-band Imaging of Interstellar Comet 3I/ATLAS from the Two-meter Twin Telescope (TTT)}",
      journal = {The Astronomer's Telegram},
         year = 2025,
        month = jul,
       volume = {17264},
        pages = {1},
       adsurl = {https://ui.adsabs.harvard.edu/abs/2025ATel17264....1A}
}

@ARTICLE{Frincke2025,
       author = {{Frincke}, Tessa T. and {Yaginuma}, Atsuhiro and {Noonan}, John W. and {Hsieh}, Henry H. and {Seligman}, Darryl Z. and {Holt}, Carrie E. and {Strader}, Jay and {Do}, Thomas and {Craig}, Peter and {Molina}, Isabella},
        title = "{Near-Discovery SOAR Photometry of the Third Interstellar Object: 3I/ATLAS}",
      journal = {\mnras},
         year = 2025,
        month = nov,
          doi = {10.1093/mnras/staf1994},
archivePrefix = {arXiv},
       eprint = {2509.02813},
 primaryClass = {astro-ph.EP},
       adsurl = {https://ui.adsabs.harvard.edu/abs/2025MNRAS.tmp.1882F}
}

@ARTICLE{Feinstein2025,
       author = {{Feinstein}, Adina D. and {Noonan}, John W. and {Seligman}, Darryl Z.},
        title = "{Precovery Observations of 3I/ATLAS from TESS Suggest Possible Distant Activity}",
      journal = {\apjl},
         year = 2025,
        month = sep,
       volume = {991},
       number = {1},
          eid = {L2},
        pages = {L2},
          doi = {10.3847/2041-8213/adfd4d},
archivePrefix = {arXiv},
       eprint = {2507.21967},
 primaryClass = {astro-ph.EP},
       adsurl = {https://ui.adsabs.harvard.edu/abs/2025ApJ...991L...2F}
}

@ARTICLE{Martinez-Palomera2025,
       author = {{Martinez-Palomera}, Jorge and {Tuson}, Amy and {Hedges}, Christina and {Dotson}, Jessie and {Barclay}, Thomas and {Powell}, Brian},
        title = "{Pre-discovery TESS Observations of Interstellar Object 3I/ATLAS}",
      journal = {arXiv e-prints},
         year = 2025,
        month = aug,
          eid = {arXiv:2508.02499},
        pages = {arXiv:2508.02499},
archivePrefix = {arXiv},
       eprint = {2508.02499},
 primaryClass = {astro-ph.EP},
       adsurl = {https://ui.adsabs.harvard.edu/abs/2025arXiv250802499M}
}

@ARTICLE{Bodewits2020,
       author = {{Bodewits}, D. and {Noonan}, J.~W. and {Feldman}, P.~D. and {Bannister}, M.~T. and {Farnocchia}, D. and {Harris}, W.~M. and {Li}, J. -Y. and {Mandt}, K.~E. and {Parker}, J. Wm. and {Xing}, Z. -X.},
        title = "{The carbon monoxide-rich interstellar comet 2I/Borisov}",
      journal = {Nature Astronomy},
         year = 2020,
        month = apr,
       volume = {4},
        pages = {867-871},
          doi = {10.1038/s41550-020-1095-2},
archivePrefix = {arXiv},
       eprint = {2004.08972},
 primaryClass = {astro-ph.EP},
       adsurl = {https://ui.adsabs.harvard.edu/abs/2020NatAs...4..867B}
}

@ARTICLE{Cordiner2020,
       author = {{Cordiner}, M.~A. and {Milam}, S.~N. and {Biver}, N. and {Bockel{\'e}e-Morvan}, D. and {Roth}, N.~X. and {Bergin}, E.~A. and {Jehin}, E. and {Remijan}, A.~J. and {Charnley}, S.~B. and {Mumma}, M.~J. and {Boissier}, J. and {Crovisier}, J. and {Paganini}, L. and {Kuan}, Y. -J. and {Lis}, D.~C.},
        title = "{Unusually high CO abundance of the first active interstellar comet}",
      journal = {Nature Astronomy},
         year = 2020,
        month = apr,
       volume = {4},
        pages = {861-866},
          doi = {10.1038/s41550-020-1087-2},
archivePrefix = {arXiv},
       eprint = {2004.09586},
 primaryClass = {astro-ph.EP},
       adsurl = {https://ui.adsabs.harvard.edu/abs/2020NatAs...4..861C}
}

@ARTICLE{Hoogendam2025,
       author = {{Hoogendam}, W.~B. and {Shappee}, B.~J. and {Wray}, J.~J. and {Yang}, B. and {Meech}, K.~J. and {Ashall}, C. and {Desai}, D.~D. and {Hart}, K. and {Hinkle}, J.~T. and {Hoffman}, A. and {Hu}, E.~M. and {Jones}, D.~O. and {Medler}, K.},
        title = "{Spatial Profiles of 3I/ATLAS CN and Ni Outgassing from Keck/KCWI Integral Field Spectroscopy}",
      journal = {arXiv e-prints},
         year = 2025,
        month = oct,
          eid = {arXiv:2510.11779},
        pages = {arXiv:2510.11779},
archivePrefix = {arXiv},
       eprint = {2510.11779},
 primaryClass = {astro-ph.EP},
       adsurl = {https://ui.adsabs.harvard.edu/abs/2025arXiv251011779H}
}

@ARTICLE{Ye2025,
       author = {{Ye}, Quanzhi and {Kelley}, Michael S.~P. and {Hsieh}, Henry H. and {Bellm}, Eric C. and {Chen}, Tracy X. and {Dekany}, Richard and {Drake}, Andrew and {Groom}, Steven L. and {Helou}, George and {Kulkarni}, Shrinivas R. and {Prince}, Thomas A. and {Riddle}, Reed},
        title = "{Prediscovery Activity of New Interstellar Object 3I/ATLAS: Rapid Brightening from 6 to 4 au}",
      journal = {\apjl},
         year = 2025,
        month = nov,
       volume = {993},
       number = {1},
          eid = {L31},
        pages = {L31},
          doi = {10.3847/2041-8213/ae147b},
archivePrefix = {arXiv},
       eprint = {2509.08792},
 primaryClass = {astro-ph.EP},
       adsurl = {https://ui.adsabs.harvard.edu/abs/2025ApJ...993L..31Y}
}

@ARTICLE{Tonry2025,
       author = {{Tonry}, John and {Denneau}, Larry and {Alarcon}, Miguel and {Clocchiatti}, Alejandro and {Erasmus}, Nicolas and {Fitzsimmons}, Alan and {Licandro}, Javier and {Meech}, Karen and {Siverd}, Robert and {Weiland}, Henry},
        title = "{ATLAS Photometry of Interstellar Object 3I/ATLAS}",
      journal = {arXiv e-prints},
         year = 2025,
        month = sep,
          eid = {arXiv:2509.05562},
        pages = {arXiv:2509.05562},
          doi = {10.48550/arXiv.2509.05562},
archivePrefix = {arXiv},
       eprint = {2509.05562},
 primaryClass = {astro-ph.EP},
       adsurl = {https://ui.adsabs.harvard.edu/abs/2025arXiv250905562T}
}

@ARTICLE{SalazarManzano2025,
       author = {{Salazar Manzano}, Luis E. and {Lin}, Hsing Wen and {Taylor}, Aster G. and {Seligman}, Darryl Z. and {Adams}, Fred C. and {Gerdes}, David W. and {Ruch}, Thomas and {Frincke}, Tessa T. and {Napier}, Kevin J.},
        title = "{Onset of CN Emission in 3I/ATLAS: Evidence for Strong Carbon-chain Depletion}",
      journal = {\apjl},
         year = 2025,
        month = nov,
       volume = {993},
       number = {1},
          eid = {L23},
        pages = {L23},
          doi = {10.3847/2041-8213/ae1232},
archivePrefix = {arXiv},
       eprint = {2509.01647},
 primaryClass = {astro-ph.EP},
       adsurl = {https://ui.adsabs.harvard.edu/abs/2025ApJ...993L..23S}
}

@ARTICLE{Lisse2025,
       author = {{Lisse}, C.~M. and {Bach}, Y.~P. and {Bryan}, S. and {Crill}, B.~P. and {Cukierman}, A. and {Dor{\'e}}, O. and {Fabinsky}, B. and {Faisst}, A. and {Korngut}, P.~M. and {Melnick}, G. and {Rustamkulov}, Z. and {Tolls}, V. and {Werner}, M. and {Sitko}, M.~L. and {Champagne}, C. and {Connelley}, M. and {Emery}, J.~P. and {Fernandez}, Y.~R. and {Yang}, B. and {the SPHEREx Science Team}},
        title = "{SPHEREx Discovery of Strong Water Ice Absorption and an Extended Carbon Dioxide Coma in 3I/ATLAS}",
      journal = {Research Notes of the American Astronomical Society},
         year = 2025,
        month = sep,
       volume = {9},
       number = {9},
          eid = {242},
        pages = {242},
          doi = {10.3847/2515-5172/ae0293},
archivePrefix = {arXiv},
       eprint = {2508.15469},
 primaryClass = {astro-ph.EP},
       adsurl = {https://ui.adsabs.harvard.edu/abs/2025RNAAS...9..242L}
}

@ARTICLE{Cordiner2025,
       author = {{Cordiner}, Martin A. and {Roth}, Nathan X. and {Kelley}, Michael S.~P. and {Bodewits}, Dennis and {Charnley}, Steven B. and {Drozdovskaya}, Maria N. and {Farnocchia}, Davide and {Micheli}, Marco and {Milam}, Stefanie N. and {Opitom}, Cyrielle and {Schwamb}, Megan E. and {Thomas}, Cristina A. and {Bagnulo}, Stefano},
        title = "{JWST Detection of a Carbon-dioxide-dominated Gas Coma Surrounding Interstellar Object 3I/ATLAS}",
      journal = {\apjl},
         year = 2025,
        month = oct,
       volume = {991},
       number = {2},
          eid = {L43},
        pages = {L43},
          doi = {10.3847/2041-8213/ae0647},
archivePrefix = {arXiv},
       eprint = {2508.18209},
 primaryClass = {astro-ph.EP},
       adsurl = {https://ui.adsabs.harvard.edu/abs/2025ApJ...991L..43C}
}

@ARTICLE{Xing2025,
       author = {{Xing}, Zexi and {Oset}, Shawn and {Noonan}, John and {Bodewits}, Dennis},
        title = "{Water Production Rates of the Interstellar Object 3I/ATLAS}",
      journal = {\apjl},
         year = 2025,
        month = oct,
       volume = {991},
       number = {2},
          eid = {L50},
        pages = {L50},
          doi = {10.3847/2041-8213/ae08ab},
archivePrefix = {arXiv},
       eprint = {2508.04675},
 primaryClass = {astro-ph.EP},
       adsurl = {https://ui.adsabs.harvard.edu/abs/2025ApJ...991L..50X}
}

@ARTICLE{Lazzarin2026,
       author = {{Lazzarin}, M. and {Mura}, A.~C. and {La Forgia}, F. and {Cremonese}, G. and {Cambianica}, P. and {Munaretto}, G. and {Farina}, A. and {Mazzotta Epifani}, E. and {Ieva}, S. and {Dotto}, E.},
        title = "{Preperihelion CN Production Rate of Interstellar Comet 3I/ATLAS}",
      journal = {\apjl},
         year = 2026,
        month = feb,
       volume = {998},
       number = {1},
          eid = {L30},
        pages = {L30},
          doi = {10.3847/2041-8213/ae3dd7},
       adsurl = {https://ui.adsabs.harvard.edu/abs/2026ApJ...998L..30L}
}

@ARTICLE{Paek2026,
       author = {{Paek}, Gregory S.~H. and {Im}, Myungshin and {Jeong}, Mankeun and {Choi}, Hyeonho and {Bach}, Yoonsoo P. and {Ishiguro}, Masateru and {Lim}, Bumhoo and {Chang}, Seo-Won and {Kim}, Ji Hoon and {Geem}, Jooyeon and {Hoogendam}, Willem B.},
        title = "{Pre-perihelion Emergence of the CN Gas Coma in 3I/ATLAS Temporally and Spatially Resolved by the 7-Dimensional Telescope}",
      journal = {arXiv e-prints},
         year = 2026,
        month = feb,
          eid = {arXiv:2602.12930},
        pages = {arXiv:2602.12930},
archivePrefix = {arXiv},
       eprint = {2602.12930},
 primaryClass = {astro-ph.EP},
       adsurl = {https://ui.adsabs.harvard.edu/abs/2026arXiv260212930P}
}

@ARTICLE{Coulson2026,
       author = {{Coulson}, Iain M. and {Kuan}, Yi-Jehng and {Charnley}, Steven B. and {Cordiner}, Martin A. and {Chuang}, Yo-Ling and {Lee}, Yueh-Ning and {Lin}, Min-Kai and {Milam}, Stefanie N. and {Pimpanuwat}, Bannawit and {Roth}, Nathan X. and {{\.Z}{\'o}{\l}towski}, Micha{\l}},
        title = "{JCMT detection of HCN emission from 3I/ATLAS at 2.1 au}",
      journal = {\mnras},
         year = 2026,
        month = feb,
       volume = {546},
       number = {2},
          eid = {stag063},
        pages = {stag063},
          doi = {10.1093/mnras/stag063},
archivePrefix = {arXiv},
       eprint = {2510.02817},
 primaryClass = {astro-ph.EP},
       adsurl = {https://ui.adsabs.harvard.edu/abs/2026MNRAS.546ag063C}
}

@ARTICLE{Roth2025,
       author = {{Roth}, Nathan X. and {Cordiner}, Martin A. and {Bockel{\'e}e-Morvan}, Dominique and {Biver}, Nicolas and {Crovisier}, Jacques and {Milam}, Stefanie N. and {Lellouch}, Emmanuel and {Santos-Sanz}, Pablo and {Lis}, Dariusz C. and {Qi}, Chunhua and {Foster}, K.~D. and {Boissier}, J{\'e}r{\'e}mie and {Furuya}, Kenji and {Moreno}, Raphael and {Charnley}, Steven B. and {Remijan}, Anthony J. and {Kuan}, Yi-Jehng and {Hart}, Lillian X.},
        title = "{CH$_3$OH and HCN in Interstellar Comet 3I/ATLAS Mapped with the ALMA Atacama Compact Array: Distinct Outgassing Behaviors and a Remarkably High CH$_3$OH/HCN Production Rate Ratio}",
      journal = {arXiv e-prints},
         year = 2025,
        month = nov,
          eid = {arXiv:2511.20845},
        pages = {arXiv:2511.20845},
          doi = {10.48550/arXiv.2511.20845},
archivePrefix = {arXiv},
       eprint = {2511.20845},
 primaryClass = {astro-ph.EP},
       adsurl = {https://ui.adsabs.harvard.edu/abs/2025arXiv251120845R}
}

@ARTICLE{Jewitt2025_NOT,
       author = {{Jewitt}, David and {Luu}, Jane},
        title = "{Pre-perihelion Development of Interstellar Comet 3I/ATLAS}",
      journal = {arXiv e-prints},
         year = 2025,
        month = oct,
          eid = {arXiv:2510.18769},
        pages = {arXiv:2510.18769},
archivePrefix = {arXiv},
       eprint = {2510.18769},
 primaryClass = {astro-ph.EP},
       adsurl = {https://ui.adsabs.harvard.edu/abs/2025arXiv251018769J}
}

@ARTICLE{Knight2017,
       author = {{Knight}, Matthew M. and {Protopapa}, Silvia and
         {Kelley}, Michael S.~P. and {Farnham}, Tony L. and {Bauer}, James M. and
         {Bodewits}, Dennis and {Feaga}, Lori M. and {Sunshine}, Jessica M.},
        title = "{On the Rotation Period and Shape of the Hyperbolic Asteroid 1I/{\textquoteleft}Oumuamua (2017 U1) from Its Lightcurve}",
      journal = {ApJL},
         year = "2017",
        month = "Dec",
       volume = {851},
       number = {2},
          eid = {L31},
        pages = {L31},
          doi = {10.3847/2041-8213/aa9d81},
archivePrefix = {arXiv},
       eprint = {1711.01402},
 primaryClass = {astro-ph.EP},
       adsurl = {https://ui.adsabs.harvard.edu/abs/2017ApJ...851L..31K}
}

@ARTICLE{Fraser2017,
       author = {{Fraser}, Wesley C. and {Pravec}, Petr and {Fitzsimmons}, Alan and
         {Lacerda}, Pedro and {Bannister}, Michele T. and {Snodgrass}, Colin and
         {Smoli{\'c}}, Igor},
        title = "{The tumbling rotational state of 1I/`Oumuamua}",
      journal = {Nature Astronomy},
         year = "2018",
        month = "Feb",
       volume = {2},
        pages = {383-386},
          doi = {10.1038/s41550-018-0398-z},
archivePrefix = {arXiv},
       eprint = {1711.11530},
 primaryClass = {astro-ph.EP},
       adsurl = {https://ui.adsabs.harvard.edu/abs/2018NatAs...2..383F}
}

@ARTICLE{Belton2018,
   author = {{Belton}, M.~J.~S. and {Hainaut}, O.~R. and {Meech}, K.~J. and 
	{Mueller}, B.~E.~A. and {Kleyna}, J.~T. and {Weaver}, H.~A. and 
	{Buie}, M.~W. and {Drahus}, M. and {Guzik}, P. and {Wainscoat}, R.~J. and 
	{Waniak}, W. and {Handzlik}, B. and {Kurowski}, S. and {Xu}, S. and 
	{Sheppard}, S.~S. and {Micheli}, M. and {Ebeling}, H. and {Keane}, J.~V.
	},
    title = "{The Excited Spin State of 1I/2017 U1 Oumuamua}",
  journal = {ApJL},
archivePrefix = "arXiv",
     year = 2018,
    month = apr,
   volume = 856,
      eid = {L21},
    pages = {L21},
      doi = {10.3847/2041-8213/aab370},
   adsurl = {https://ui.adsabs.harvard.edu/abs/2018ApJ...856L..21B}
}

@ARTICLE{Mashchenko2019,
       author = {{Mashchenko}, Sergey},
        title = "{Modelling the light curve of `Oumuamua: evidence for torque and disc-like shape}",
      journal = {MNRAS},
         year = "2019",
        month = "Nov",
       volume = {489},
       number = {3},
        pages = {3003-3021},
          doi = {10.1093/mnras/stz2380},
archivePrefix = {arXiv},
       eprint = {1906.03696},
 primaryClass = {astro-ph.EP},
       adsurl = {https://ui.adsabs.harvard.edu/abs/2019MNRAS.489.3003M}
}

@ARTICLE{Fitzsimmons2017,
       author = {{Fitzsimmons}, Alan and {Snodgrass}, Colin and {Rozitis}, Ben and
         {Yang}, Bin and {Hyland}, M{\'e}abh and {Seccull}, Tom and
         {Bannister}, Michele T. and {Fraser}, Wesley C. and {Jedicke}, Robert and
         {Lacerda}, Pedro},
        title = "{Spectroscopy and thermal modelling of the first interstellar object 1I/2017 U1 `Oumuamua}",
      journal = {Nature Astronomy},
         year = "2018",
        month = "Dec",
       volume = {2},
        pages = {133-137},
          doi = {10.1038/s41550-017-0361-4},
archivePrefix = {arXiv},
       eprint = {1712.06552},
 primaryClass = {astro-ph.EP},
       adsurl = {https://ui.adsabs.harvard.edu/abs/2018NatAs...2..133F}
}

@ARTICLE{Bannister2017,
       author = {{Bannister}, Michele T. and {Schwamb}, Megan E. and {Fraser}, Wesley C. and
         {Marsset}, Michael and {Fitzsimmons}, Alan and {Benecchi}, Susan D. and
         {Lacerda}, Pedro and {Pike}, Rosemary E. and {Kavelaars}, J.~J. and
         {Smith}, Adam B. and {Stewart}, Sunny O. and {Wang}, Shiang-Yu and
         {Lehner}, Matthew J.},
        title = "{Col-OSSOS: Colors of the Interstellar Planetesimal 1I/{\textquoteleft}Oumuamua}",
      journal = {ApJL},
         year = "2017",
        month = "Dec",
       volume = {851},
       number = {2},
          eid = {L38},
        pages = {L38},
          doi = {10.3847/2041-8213/aaa07c},
archivePrefix = {arXiv},
       eprint = {1711.06214},
 primaryClass = {astro-ph.EP},
       adsurl = {https://ui.adsabs.harvard.edu/abs/2017ApJ...851L..38B}
}

@ARTICLE{Taylor2023,
       author = {{Taylor}, Aster G. and {Seligman}, Darryl Z. and {Hainaut}, Olivier R. and {Meech}, Karen J.},
        title = "{Fitting the Light Curve of 1I/'Oumuamua with a Nonprincipal Axis Rotational Model and Outgassing Torques}",
      journal = {\psj},
         year = 2023,
        month = oct,
       volume = {4},
       number = {10},
          eid = {186},
        pages = {186},
          doi = {10.3847/PSJ/acf617},
archivePrefix = {arXiv},
       eprint = {2309.01820},
 primaryClass = {astro-ph.EP},
       adsurl = {https://ui.adsabs.harvard.edu/abs/2023PSJ.....4..186T}
}

@ARTICLE{Jewitt2023ARAA,
       author = {{Jewitt}, David and {Seligman}, Darryl Z.},
        title = "{The Interstellar Interlopers}",
      journal = {\araa},
         year = 2023,
        month = aug,
       volume = {61},
        pages = {197-236},
          doi = {10.1146/annurev-astro-071221-054221},
archivePrefix = {arXiv},
       eprint = {2209.08182},
 primaryClass = {astro-ph.EP},
       adsurl = {https://ui.adsabs.harvard.edu/abs/2023ARA&A..61..197J}
}

@ARTICLE{MoroMartin2022,
       author = {{Moro-Mart{\'\i}n}, Amaya},
        title = "{Interstellar planetesimals}",
      journal = {arXiv e-prints},
         year = 2022,
        month = may,
          eid = {arXiv:2205.04277},
        pages = {arXiv:2205.04277},
          doi = {10.48550/arXiv.2205.04277},
archivePrefix = {arXiv},
       eprint = {2205.04277},
 primaryClass = {astro-ph.EP},
       adsurl = {https://ui.adsabs.harvard.edu/abs/2022arXiv220504277M}
}

@INCOLLECTION{Fitzsimmons2024,
       author = {{Fitzsimmons}, Alan and {Meech}, Karen and {Matr{\`a}}, Luca and {Pfalzner}, Susanne},
        title = "{Interstellar Objects and Exocomets}",
    booktitle = {Comets III},
         year = 2024,
       editor = {{Meech}, Karen. J. and {Combi}, Michael. R. and {Bockel{\'e}e-Morvan}, Dominique and {Raymodn}, Sean. N. and {Zolensky}, Michael. E.},
        pages = {731-766},
       adsurl = {https://ui.adsabs.harvard.edu/abs/2024come.book..731F}
}

@ARTICLE{Seligman2023,
       author = {{Seligman}, Darryl Z. and {Moro-Mart{\'\i}n}, Amaya},
        title = "{Interstellar objects}",
      journal = {Contemporary Physics},
         year = 2023,
        month = jul,
       volume = {63},
       number = {3},
        pages = {200-232},
          doi = {10.1080/00107514.2023.2203976},
archivePrefix = {arXiv},
       eprint = {2304.00568},
 primaryClass = {astro-ph.EP},
       adsurl = {https://ui.adsabs.harvard.edu/abs/2022ConPh..63..200S}
}

@ARTICLE{Taylor2025,
       author = {{Taylor}, Aster G. and {Seligman}, Darryl Z.},
        title = "{The Kinematic Age of 3I/ATLAS and Its Implications for Early Planet Formation}",
      journal = {\apjl},
         year = 2025,
        month = sep,
       volume = {990},
       number = {1},
          eid = {L14},
        pages = {L14},
          doi = {10.3847/2041-8213/adfa28},
archivePrefix = {arXiv},
       eprint = {2507.08111},
 primaryClass = {astro-ph.EP},
       adsurl = {https://ui.adsabs.harvard.edu/abs/2025ApJ...990L..14T}
}

@ARTICLE{Mamajek2017,
   author = {{Mamajek}, E.},
    title = "{Kinematics of the Interstellar Vagabond 1I/Oumuamua (A/2017 U1)}",
  journal = {Research Notes of the American Astronomical Society},
archivePrefix = "arXiv",
   eprint = {1710.11364},
 primaryClass = "astro-ph.EP",
     year = 2017,
    month = dec,
   volume = 1,
   number = 1,
      eid = {21},
    pages = {21},
      doi = {10.3847/2515-5172/aa9bdc},
   adsurl = {http://adsabs.harvard.edu/abs/2017RNAAS...1a..21M}
}

\appendix
\onecolumn
\section{Expanded Equations}

In section \ref{subsec: 3D_skymotion}, we lay out the method for applying the generalized vector form for plane-of-sky motion from Equation \ref{eqtn: skymotion_vectoreqtn} to an object on 3D orbit as a function of orbital elements. We start by defining the heliocentric distance and velocity vectors in the orbital plane of the object, shown in Equations \ref{eqtn: ISO_position_perifocal} and \ref{eqtn: v_xy_elements}, and rotate into the reference frame of the ecliptic plane of the Sun, with the $x^\prime$ axis defined along the vernal equinox. The three rotation matrices needed for this coordinate transformation are defined in Equations \ref{eqtn: Rot1_matrix_w}, \ref{eqtn: Rot2_matrix_i}, and \ref{eqtn: Rot3_matrix_Om}. Combining these equations as shown in Equation \ref{eqtn: Full_Rotation_Matrix_unsimplified}, we obtain a full rotation matrix,

\begin{equation}
\overleftrightarrow{{M}}(\omega, i, \Omega) =
\begin{pmatrix}
\cos{\omega}\,\cos{\Omega}-\cos{i}\,\sin{\omega}\,\sin{\Omega} & \sin{\omega}\,\cos{\Omega}-\cos{i}\,\cos{\omega}\,\sin{\Omega} & \sin{i}\,\sin{\Omega}\\
-\cos{i}\,\sin{\omega}\,\cos{\Omega}-\cos{\omega}\,\sin{\Omega} & \cos{i}\,\cos{\omega}\,\cos{\Omega}-\sin{\omega}\,\sin{\Omega} & \sin{i}\,\cos{\Omega}\\
\sin{i}\,\sin{\omega} & -\cos{\omega}\,\sin{i} & \cos{i}
\end{pmatrix} \ .
\label{eqtn: Full_Rotation_Matrix}
\end{equation}

We apply this coordinate transformation matrix to the heliocentric distance and velocity vectors of the object from Equations \ref{eqtn: ISO_position_perifocal} and \ref{eqtn: v_xy_elements}, and obtain these vectors in our new coordinate basis in the ecliptic plane, as shown in Equations \ref{eqtn: rvec_prime3D} and \ref{eqtn: vvec_prime3D}. Next, we find the distance vector between the earth and the object by subtracting Equations \ref{eqtn: rearthvec_prime3D} and \ref{eqtn: rearthvec_prime3D} as in Equation \ref{eqtn: d_separation}. We plug in this distance vector, as well as the object velocity vector in the ecliptic coordinate frame from Equation \ref{eqtn: vvec_prime3D} into the generalized vector formula for sky motion from Equation \ref{eqtn: skymotion_vectoreqtn}. After further simplification, the full equation for the apparent sky motion of an object on a 3D orbit as viewed from Earth is given by:

\begin{equation}
\begin{split}
\frac{d\theta}{dt} &= 
\sqrt{\frac{GM}{\left| a(1-e^2) \right|}} 
\frac{1}{4} 
\bigg| \frac{1}{r^2 + r_\oplus^2 - 2 r r_\oplus 
[\cos(f-\omega)\cos(f_\oplus+\Omega+\omega_\oplus) + \cos{i}\,\sin(f-\omega)\sin(f_\oplus+\Omega+\omega_\oplus)]} \bigg| \\
&\quad \times 
\Big[
8(2+e^2)r^2+10(1+e^2)r_\oplus^2 + 4e(8r^2+5r_\oplus^2)\cos{f}+8e^2r^2\cos{2f} \\
&\quad + \frac{1}{2} r_\oplus \Big(-4er_\oplus \cos(f-2i) -4(1+e^2)r_\oplus \cos{2i} - 4e r_\oplus \cos(f+2i) + 8e r_\oplus \cos(f- 2\omega) + 4r_\oplus \cos(2f - 2\omega) \\
&\quad - 2r_\oplus \cos(2f - 2i - 2\omega) - 2 e^2 r_\oplus \cos(2i - 2\omega) - 4 e r_\oplus \cos(f + 2i - 2\omega) - 2 r_\oplus \cos(2(f + i - \omega)) \\
&\quad + 4 e^2 r_\oplus \cos 2\omega - 2 e^2 r_\oplus \cos 2(i + \omega) - 4 e r_\oplus \cos(f - 2(i + \omega)) + 6 r_\oplus \cos(2 f - 2 f_\oplus - 2\omega - 2\Omega - 2\omega_\oplus) \\
&\quad + r_\oplus \cos(2 f - 2 f_\oplus - 2 i - 2 \omega - 2 \Omega - 2 \omega_\oplus) - 32 r \cos(f - f_\oplus - \omega - \Omega - \omega_\oplus) - 16 e^2 r \cos(f - f_\oplus - \omega - \Omega - \omega_\oplus) \\
&\quad - 16 e r \cos(2f - f_\oplus - \omega - \Omega - \omega_\oplus) - 16 r \cos(f - f_\oplus - i - \omega - \Omega - \omega_\oplus) - 8 e^2 r \cos(f - f_\oplus - i - \omega - \Omega - \omega_\oplus) \\
&\quad - 8 e r \cos(2f - f_\oplus - i - \omega - \Omega - \omega_\oplus) - 16 r \cos(f - f_\oplus + i - \omega - \Omega - \omega_\oplus) - 8 e^2 r \cos(f - f_\oplus + i - \omega - \Omega - \omega_\oplus) \\
&\quad + r_\oplus \cos(2(f - f_\oplus + i - \omega - \Omega - \omega_\oplus)) - 8 e r \cos(2f - f_\oplus + i - \omega - \Omega - \omega_\oplus) - 16 e^2 r \cos(f - f_\oplus + \omega - \Omega - \omega_\oplus) \\
&\quad + 8 e^2 r \cos(f - f_\oplus - i + \omega - \Omega - \omega_\oplus) + 8 e^2 r \cos(f - f_\oplus + i + \omega - \Omega - \omega_\oplus) - 4 r_\oplus \cos(2(f_\oplus + \Omega + \omega_\oplus)) \\
&\quad - 4 e^2 r_\oplus \cos(2(f_\oplus + \Omega + \omega_\oplus)) + 2 r_\oplus \cos(2(f_\oplus - i + \Omega + \omega_\oplus)) + 2 e^2 r_\oplus \cos(2(f_\oplus - i + \Omega + \omega_\oplus)) \\
&\quad  + 2 r_\oplus \cos(2(f_\oplus + i + \Omega + \omega_\oplus)) + 2 e^2 r_\oplus \cos(2(f_\oplus + i + \Omega + \omega_\oplus)) - 48 e r \cos(f_\oplus - \omega + \Omega + \omega_\oplus) \\
&\quad + 6 e^2 r_\oplus \cos(2(f_\oplus - \omega + \Omega + \omega_\oplus)) - 32 r \cos(f + f_\oplus - \omega + \Omega + \omega_\oplus) - 16 e^2 r \cos(f + f_\oplus - \omega + \Omega + \omega_\oplus) \\
&\quad + 6 r_\oplus \cos(2(f + f_\oplus - \omega + \Omega + \omega_\oplus)) - 16 e r \cos(2f + f_\oplus - \omega + \Omega + \omega_\oplus) + 24 e r \cos(f_\oplus - i - \omega + \Omega + \omega_\oplus) \\
&\quad + e^2 r_\oplus \cos(2(f_\oplus - i - \omega + \Omega + \omega_\oplus)) + 16 r \cos(f + f_\oplus - i - \omega + \Omega + \omega_\oplus) + 8 e^2 r \cos(f + f_\oplus - i - \omega + \Omega + \omega_\oplus) \\
&\quad + r_\oplus \cos(2(f + f_\oplus - i - \omega + \Omega + \omega_\oplus)) + 8 e r \cos(2f + f_\oplus - i - \omega + \Omega + \omega_\oplus) + 24 e r \cos(f_\oplus + i - \omega + \Omega + \omega_\oplus) \\
&\quad + e^2 r_\oplus \cos(2(f_\oplus + i - \omega + \Omega + \omega_\oplus)) + 16 r \cos(f + f_\oplus + i - \omega + \Omega + \omega_\oplus) + 8 e^2 r \cos(f + f_\oplus + i - \omega + \Omega + \omega_\oplus) \\
&\quad + r_\oplus \cos(2(f + f_\oplus + i - \omega + \Omega + \omega_\oplus)) + 8 e r \cos(2f + f_\oplus + i - \omega + \Omega + \omega_\oplus) - 48 e r \cos(f_\oplus + \omega + \Omega + \omega_\oplus) \\
&\quad + 6 e^2 r_\oplus \cos(2(f_\oplus + \omega + \Omega + \omega_\oplus)) - 16 e^2 r \cos(f + f_\oplus + \omega + \Omega + \omega_\oplus) - 24 e r \cos(f_\oplus - i + \omega + \Omega + \omega_\oplus) \\
&\quad + e^2 r_\oplus \cos(2(f_\oplus - i + \omega + \Omega + \omega_\oplus)) - 8 e^2 r \cos(f + f_\oplus - i + \omega + \Omega + \omega_\oplus) - 24 e r \cos(f_\oplus + i + \omega + \Omega + \omega_\oplus) \\
&\quad + e^2 r_\oplus \cos(2(f_\oplus + i + \omega + \Omega + \omega_\oplus)) - 8 e^2 r \cos(f + f_\oplus + i + \omega + \Omega + \omega_\oplus) - 4 e r_\oplus \cos(f - 2(f_\oplus + \Omega + \omega_\oplus)) \\
&\quad - 4 e r_\oplus \cos(f + 2(f_\oplus + \Omega + \omega_\oplus)) + 2 e r_\oplus \cos(f - 2(f_\oplus - i + \Omega + \omega_\oplus)) + 2 e r_\oplus \cos(f + 2(f_\oplus - i + \Omega + \omega_\oplus)) \\
&\quad + 2 e r_\oplus \cos(f - 2(f_\oplus + i + \Omega + \omega_\oplus)) + 2 e r_\oplus \cos(f + 2(f_\oplus + i + \Omega + \omega_\oplus)) - 4 e^2 r_\oplus \cos(2 f_\oplus - i + 2(-\omega + \Omega + \omega_\oplus)) \\
&\quad - 8 e r_\oplus \cos(f + 2 f_\oplus - i + 2(-\omega + \Omega + \omega_\oplus)) - 4 r_\oplus \cos(2f + 2 f_\oplus - i + 2(-\omega + \Omega + \omega_\oplus)) - 4 e^2 r_\oplus \cos(2 f_\oplus + i + 2(-\omega + \Omega + \omega_\oplus)) \\
&\quad - 8 e r_\oplus \cos(f + 2 f_\oplus + i + 2(-\omega + \Omega + \omega_\oplus)) - 4 r_\oplus \cos(2 f + 2 f_\oplus + i + 2(-\omega + \Omega + \omega_\oplus)) + 12 e r_\oplus \cos(f + 2(f_\oplus - \omega + \Omega + \omega_\oplus)) \\
&\quad + 2 e r_\oplus \cos(f + 2(f_\oplus - i - \omega + \Omega + \omega_\oplus)) + 2 e r_\oplus \cos(f + 2(f_\oplus + i - \omega + \Omega + \omega_\oplus)) + 8 e r_\oplus \cos(f - 2 f_\oplus - i - 2(\omega + \Omega + \omega_\oplus)) \\
&\quad + 4 r_\oplus \cos(2f - 2 f_\oplus - i - 2(\omega + \Omega + \omega_\oplus)) + 8 e r_\oplus \cos(f - 2 f_\oplus + i - 2(\omega + \Omega + \omega_\oplus)) + 4 r_\oplus \cos(2f - 2 f_\oplus + i - 2(\omega + \Omega + \omega_\oplus)) \\
&\quad + 4 e^2 r_\oplus \cos(2 f_\oplus - i + 2(\omega + \Omega + \omega_\oplus)) + 4 e^2 r_\oplus \cos(2 f_\oplus + i + 2(\omega + \Omega + \omega_\oplus)) + 12 e r_\oplus \cos(f - 2(f_\oplus + \omega + \Omega + \omega_\oplus))  \\
&\quad + 2 e r_\oplus \cos(f - 2(f_\oplus - i + \omega + \Omega + \omega_\oplus)) + 2 e r_\oplus \cos(f - 2(f_\oplus + i + \omega + \Omega + \omega_\oplus))
\Big)
\Big]^\frac{1}{2}
\end{split}
\label{eqtn:3D_skymotion_orbparams_full}
\end{equation}

\section{Comet-like Object Values}

We present additional median sky motion values of the sky motion distributions for comet-like objects in Figure \ref{fig: brightening_limit_motions}, as well as a supplementary summary statistics figure.

\begin{table}
\begin{center}
\begin{tabular}{ccc}
\hline
\multicolumn{1}{c}{Magnitude} & \multicolumn{1}{c}{Median $d\theta/dt$ [$^\circ$/d]} & \multicolumn{1}{c}{Median $d\theta/dt$ [$^\circ$/d]}\\
\multicolumn{1}{c}{ $m_V$ } & \multicolumn{1}{c}{Small ($R_c=100\,m$)} & \multicolumn{1}{c}{Large ($R_c=1\,km$)}\\
\hline
\underline{1.2 AU}& --- & ---\\
17.0 & {1.28} & {0.21} \\
19.0 & {0.89} & {0.12} \\
21.0 & {0.55} & {0.07} \\
23.0 & {0.33} & {0.04} \\
25.0 & {0.19} & {0.03} \\
27.0 & {0.11} & {0.02} \\
\hline
\underline{3.0 AU}& --- & ---\\
17.0 & {1.14} & {0.21} \\
19.0 & {0.79} & {0.12} \\
21.0 & {0.54} & {0.07} \\
23.0 & {0.33} & {0.04} \\
25.0 & {0.19} & {0.02} \\
27.0 & {0.11} & {0.01} \\
\hline
\underline{5.0 AU}& --- & ---\\
17.0 & {1.12} & {0.23} \\
19.0 & {0.78} & {0.13} \\
21.0 & {0.51} & {0.07} \\
23.0 & {0.33} & {0.04} \\
25.0 & {0.21} & {0.02} \\
27.0 & {0.12} & {0.01} \\
\hline
\end{tabular}
\caption{Median sky motions of comet-like interstellar objects in 1.2 AU, 3.0 AU, and 5.0 AU spheres at limiting {apparent v-band} magnitudes from the distributions in Figure \ref{fig: brightening_limit_motions}.}
\label{table:brightening_median_motions} 
\end{center}
\end{table}

\begin{figure}
    \centering
    \includegraphics[width=0.45\linewidth]{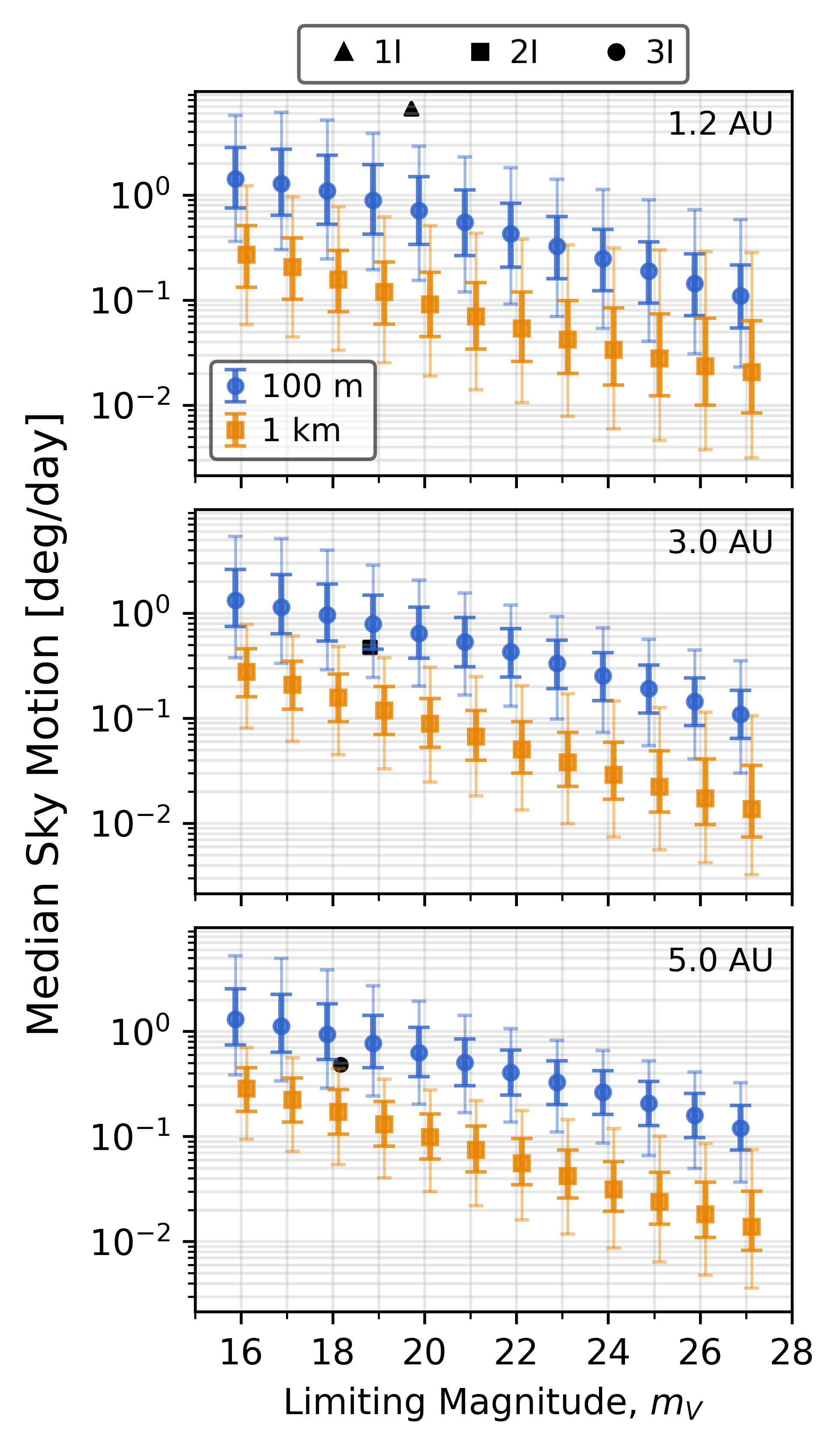}
    \caption{Summary statistics showing the median, 1-sigma, and 2-sigma quantiles for distributions of sky motions for the comet-like populations shown in Figure \ref{fig: brightening_limit_motions}. Points represent the median sky motion for a given distribution; error bars represent 1-sigma and 2-sigma quantiles, respectively. A sample of specific median values are provided in Table \ref{table:brightening_median_motions}. Reference points for the discovery conditions of the three known interstellar objects are plotted using the values in Table \ref{table: known_ISO_values_discovery_max}.}
    \label{fig:smlg_stats}
\end{figure}

\bsp	
\label{lastpage}
\end{document}